\documentclass[twocolumn]{aastex631}

\shortauthors{Viaña et al.}

\usepackage[utf8]{inputenc}
\usepackage{verbatim}
\usepackage{lmodern} 
\usepackage{booktabs}  
\usepackage{soul}
\usepackage{ulem}

\begin{document}

\title{\large \textbf{\texttt{LensNet}}: Enhancing Real-time Microlensing Event Discovery with Recurrent Neural Networks in the Korea Microlensing Telescope Network}




\correspondingauthor{Javier Viaña}
\email{vianajr@mit.edu}

\author[0000-0002-0563-784X]{Javier Viaña}
\altaffiliation{Mauricio and Carlota Botton Fellow at MIT}
\affiliation{Department of Physics, Massachusetts Institute of Technology, Cambridge, MA 02139, USA}
\affiliation{Kavli Institute for Astrophysics and Space Research, Massachusetts Institute of Technology, Cambridge, MA 02139, USA}

\author[0000-0002-9241-4117]{Kyu-Ha Hwang}
\affiliation{Korea Astronomy and Space Science Institute, Daejeon 34055, Republic of Korea}
\email{kyuha@kasi.re.kr}

\author[0000-0002-7564-6047]{Zo\"e de Beurs}
\affiliation{Department of Physics, Massachusetts Institute of Technology, Cambridge, MA 02139, USA}
\affiliation{Department of Earth, Atmospheric and Planetary Sciences, Massachusetts Institute of Technology, Cambridge, MA 02139, USA}
\affiliation{Kavli Institute for Astrophysics and Space Research, Massachusetts Institute of Technology, Cambridge, MA 02139, USA}

\author[0000-0001-9481-7123]{Jennifer C. Yee}
\affiliation{Center for Astrophysics $|$ Harvard \& Smithsonian, 60 Garden St.,Cambridge, MA 02138, USA}
\email{jyee@cfa.harvard.edu}

\author[0000-0001-7246-5438]{Andrew Vanderburg}
\affiliation{Department of Physics, Massachusetts Institute of Technology, Cambridge, MA 02139, USA}
\affiliation{Kavli Institute for Astrophysics and Space Research, Massachusetts Institute of Technology, Cambridge, MA 02139, USA}


\author[0000-0003-3316-4012]{Michael D. Albrow}
\affiliation{University of Canterbury, School of Physical and Chemical Sciences, Private Bag 4800, Christchurch 8020, New Zealand}

\author[0000-0001-6285-4528]{Sun-Ju Chung}
\affiliation{Korea Astronomy and Space Science Institute, Daejeon 34055, Republic of Korea}

\author{Andrew Gould} 
\affiliation{Max-Planck-Institute for Astronomy, K\"onigstuhl 17, 69117 Heidelberg, Germany}
\affiliation{Department of Astronomy, Ohio State University, 140 W. 18th Ave., Columbus, OH 43210, USA}

\author[0000-0002-2641-9964]{Cheongho Han}
\affiliation{Department of Physics, Chungbuk National University, Cheongju 28644, Republic of Korea}

\author[0000-0002-0314-6000]{Youn Kil Jung}
\affiliation{Korea Astronomy and Space Science Institute, Daejeon 34055, Republic of Korea}
\affiliation{National University of Science and Technology (UST), Daejeon 34113, Republic of Korea}

\author[0000-0001-9823-2907]{Yoon-Hyun Ryu} 
\affiliation{Korea Astronomy and Space Science Institute, Daejeon 34055, Republic of Korea}

\author[0000-0002-4355-9838]{In-Gu Shin}
\affiliation{Center for Astrophysics $|$ Harvard \& Smithsonian, 60 Garden St.,Cambridge, MA 02138, USA}

\author[0000-0003-1525-5041]{Yossi Shvartzvald}
\affiliation{Department of Particle Physics and Astrophysics, Weizmann Institute of Science, Rehovot 7610001, Israel}

\author[0000-0003-0626-8465]{Hongjing Yang}
\affiliation{Department of Astronomy, Tsinghua University, Beijing 100084, China}

\author[0000-0001-6000-3463]{Weicheng Zang}
\affiliation{Center for Astrophysics $|$ Harvard \& Smithsonian, 60 Garden St.,Cambridge, MA 02138, USA}


\author[0000-0002-7511-2950]{Sang-Mok Cha} 
\affiliation{Korea Astronomy and Space Science Institute, Daejeon 34055, Republic of Korea}
\affiliation{School of Space Research, Kyung Hee University, Yongin, Kyeonggi 17104, Republic of Korea} 

\author{Dong-Jin Kim}
\affiliation{Korea Astronomy and Space Science Institute, Daejeon 34055, Republic of Korea}

\author[0000-0003-0562-5643]{Seung-Lee Kim} 
\affiliation{Korea Astronomy and Space Science Institute, Daejeon 34055, Republic of Korea}

\author[0000-0003-0043-3925]{Chung-Uk Lee}
\affiliation{Korea Astronomy and Space Science Institute, Daejeon 34055, Republic of Korea}

\author[0009-0000-5737-0908]{Dong-Joo Lee} 
\affiliation{Korea Astronomy and Space Science Institute, Daejeon 34055, Republic of Korea}

\author[0000-0001-7594-8072]{Yongseok Lee} 
\affiliation{Korea Astronomy and Space Science Institute, Daejeon 34055, Republic of Korea}
\affiliation{School of Space Research, Kyung Hee University, Yongin, Kyeonggi 17104, Republic of Korea}

\author[0000-0002-6982-7722]{Byeong-Gon Park}
\affiliation{Korea Astronomy and Space Science Institute, Daejeon 34055, Republic of Korea}

\author[0000-0003-1435-3053]{Richard W. Pogge} 
\affiliation{Department of Astronomy, Ohio State University, 140 West 18th Ave., Columbus, OH  43210, USA}
\affiliation{Center for Cosmology and AstroParticle Physics, Ohio State University, 191 West Woodruff Ave., Columbus, OH 43210, USA}




\begin{abstract}

Traditional microlensing event vetting methods require highly trained human experts, and the process is both complex and time-consuming. This reliance on manual inspection often leads to inefficiencies and constrains the ability to scale for widespread exoplanet detection, ultimately hindering discovery rates. To address the limits of traditional microlensing event vetting, we have developed LensNet, a machine learning pipeline specifically designed to distinguish legitimate microlensing events from false positives caused by instrumental artifacts, such as pixel bleed trails and diffraction spikes. Our system operates in conjunction with a preliminary algorithm that detects increasing trends in flux. These flagged instances are then passed to LensNet for further classification, allowing for timely alerts and follow-up observations. Tailored for the multi-observatory setup of the Korea Microlensing Telescope Network (KMTNet) and trained on a rich dataset of manually classified events, LensNet is optimized for early detection and warning of microlensing occurrences, enabling astronomers to organize follow-up observations promptly. The internal model of the pipeline employs a multi-branch Recurrent Neural Network (RNN) architecture that evaluates time-series flux data with contextual information, including sky background, the full width at half maximum of the target star, flux errors, PSF quality flags, and air mass for each observation. We demonstrate a classification accuracy above 87.5\%, and anticipate further improvements as we expand our training set and continue to refine the algorithm.

\end{abstract}






\section{Introduction} \label{sec:intro}

\begin{figure*}[h]
    \centering
    \includegraphics[width=\textwidth]{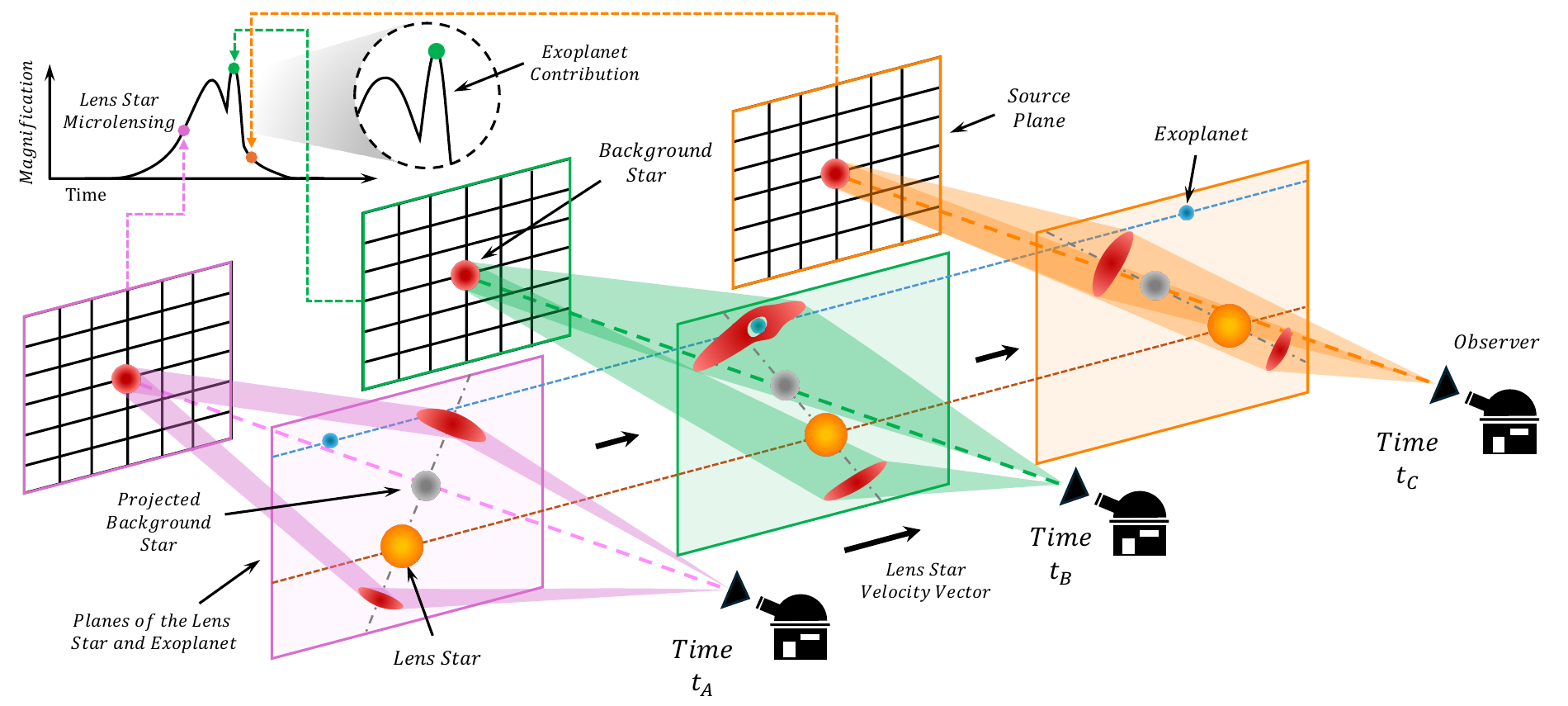}
    \caption{Illustration of the gravitational microlensing phenomenon used to detect exoplanets. As a foreground lens star passes in front of a more distant background star, the gravitational field of the lens star warps spacetime, magnifying the light from the background star. The diagram shows three stages of this event, where the lens star and its exoplanet's relative positions cause distinct distortions in the background star's light, as depicted by the warped grid lines. The resulting light curve, displayed in the inset, features two peaks: a primary one due to the lens star and a smaller secondary peak caused by the exoplanet.}
    \label{fig:fig_1}
\end{figure*}

Microlensing offers a powerful and distinctive approach to exoplanet detection by leveraging a planet's gravitational perturbation of light from distant sources. Unlike other methods, microlensing excels at uncovering exoplanets in intermediate orbits (1-5 AU) around stars that are often out of reach for radial velocity and transit techniques. Its sensitivity to planets in a wider range of orbits, including those in higher-inclination systems, makes it an invaluable tool for expanding our understanding of planetary systems.  This method serves as a crucial complement to other techniques, filling gaps in planetary discovery by revealing worlds that would otherwise remain hidden from view. This unique capability significantly enhances the breadth of exoplanet exploration, enabling the detection of a broader diversity of planetary systems. Fig. \ref{fig:fig_1} provides an illustration of the microlensing phenomenon, highlighting how the gravitational field of a lens star and its exoplanet distort the light from a background star, leading to a characteristic Paczy\'{n}ski light curve \citep{Paczynski86b}.

The field of microlensing has evolved significantly since the 1990s, when surveys of the Galactic bulge typically found a few dozen gravitational microlensing events per year \citep{Alcock1996, Udalski1994}. First, as the focus shifted to exoplanets, a two-tiered strategy \citep{GouldLoeb92} developed to combine real-time event detection in wide-field, low-cadence (few observations per day or week) surveys  \citep{Udalski94_EWS,Bond01} with targeted follow-up (multiple observations per hour) to find and characterize exoplanet signals \citep[e.g.,][]{Bond04,Udalski05,Beaulieu06}. Then, as wide-format cameras continued to grow in size, it became possible to achieve a high enough cadence to routinely detect planets in survey data, without follow-up observations \citep[e.g.,][]{Shvartzvald16}. 

The Korea Microlensing Telescope Network \citep[KMTNet;][]{Kim2016} provides valuable high-sensitivity capabilities to advance ground-based microlensing observations. KMTNet consists of three 1.6-meter telescopes located at strategic sites across the globe—Cerro Tololo Inter-American Observatory (CTIO) in Chile, South African Astronomical Observatory (SAAO) in South Africa, and Siding Spring Observatory (SSO) in Australia. This global distribution enables nearly continuous coverage of the Galactic bulge, allowing KMTNet to capture microlensing events as they occur in real time. Each telescope is equipped with a large field-of-view camera that covers 4 square degrees of the sky. In total, the survey monitors approximately 500 million stars nightly, ensuring comprehensive and efficient detection of microlensing events.

KMTNet operates with a tiered cadence strategy that optimizes its near-continuous coverage of the Galactic bulge. Six key fields (see Fig. \ref{fig:fig_2_BLG_fields}, which is a reproduction of \citealt{KimKim18_EF}) are monitored at a high cadence of 0.5 hours, with overlapping field pairs (01/41, 02/42, 03/43) receiving four observations per hour. A subset of stars in the overlap between 02/42 and 03/43 benefits from an even higher cadence of eight observations per hour. Other fields are observed at slower rates of 1, 0.4, or 0.2 observations per hour, depending on their intrinsic event rates. There are also other, rare cases, of overlap between neighboring fields. 

While these cadences are often high enough to detect and characterize planetary signals in events across the Galactic Bulge (e.g., papers in the ``Systematic KMTNet Planetary Anomaly Search" series, starting with \citealt{Zang21AF1}), there continues to be value in additional follow-up observations of known microlensing events. High-magnification events can still be targeted for follow-up observations and that interest has been extended into the ``moderate" magnification (peak magnification $A_{\rm peak} > 20$) regime \citep[e.g., ][]{Abe13, Yee21_ob0960, Zang21_kb0414}. Real-time alerts also allow for astrometric \citep{Sahu22_BH, Lam22_BH, Mroz22} or interferometric \citep{Dong19_GRAVITY, Zang20Kojima} observations to characterize the lenses (which is important for black hole searches in addition to planets) or spectroscopic observations to characterize the source stars \citep[e.g.,][]{Bensby13}.

Furthermore, the advent of low-cadence, all-sky surveys (e.g., ASAS-SN \citealt{Shappee14_ASAS, Kochanek17_ASAS}, Zwicky Transient Facility \citealt{Bellm19_ZTF,Graham19_ZTF, Masci19_ZTF}, etc.) have created new opportunities for detecting microlensing events and potentially following them up with high-cadence observations to detect planets \citep[e.g.][]{Nucita18}. Notable among these is Gaia, which has a real-time alert system for identifying microlensing events that has led to publication of several events with follow-up observations \citep[e.g.,][]{Wyrzykowski20_Gaia16aye,Rybicki22_Gaia19bld}. In addition, the upcoming Rubin Legacy Survey of Space and Time \citep[LSST; ][]{LSST09} is capable of detecting microlensing events \citep{Gould13LSST, LSST17, Street23_LSST, Abrams23_LSST}, so there has been interest in developing the capacity to identify microlensing events in real-time \citep{Godines19}.

While real-time detection of microlensing events continues to be relevant, it remains a difficult problem and many existing solutions require labor-intensive by-eye reviews. Real-time alert detection differs significantly from post-season event detection, because the characteristic features that define a microlensing light curve (e.g., the Paczy\'{n}ski shape or caustic crossings) may not be apparent, yet. In fact, the goal is usually to alert the events as early as they can be reliably identified to maximize the potential for follow-up observations. Under these circumstances, the light curves are roughly described as ``smoothly increasing in brightness above the baseline level," which is vague enough that it applies to a wide variety of astrophysical phenomena and even certain types of correlated noise and may be further complicated by gaps in the data due to observability constraints such as weather. 

The difficulty of this task also highlights the potential to apply modern machine learning (ML) techniques, which have already been shown to be very efficient in improving early detections \citep{Gezer22}. For example, \citet{Godines19} developed a Random Forest classifier for identifying ongoing microlensing events in real-time from a simulated LSST data stream, and tested the performance of the algorithm on data from OGLE-II, PTF, and ZTF. ML algorithms have also been applied to post-season event detection in real data \citep{Wyrzykowski15_OGLEIII, Chu19, Mroz20_NN} and in simulated LSST data \citep[e.g.,][]{Boone19_Avocado}. However, as we have discussed, there is significantly more information available in post-season event detection than real-time detection, making these fundamentally different applications.

In this work, we focus on the real-time ``alert" identification process for the Korea Microlensing Telescope Network (KMTNet) survey. This dataset differs significantly from prior survey datasets mined for events in real-time in that it combines data from three different sites and achieves a very high cadence of observations (typically, $\Gamma =$4--0.2 obs hr$^{-1}$), sometimes with overlap between the sites. The existing KMTNet AlertFinder algorithm \citep{Kim18_AF} has been in operation since 2018 and even operated in 2020 during the COVID-19 pandemic (albeit on a reduced datastream due to observatory closures). The first season included only the Northern Bulge fields, but the scope was expanded to the full KMTNet survey area in 2019. 

\begin{figure}[h!]
    \centering
    \includegraphics[width=\columnwidth]{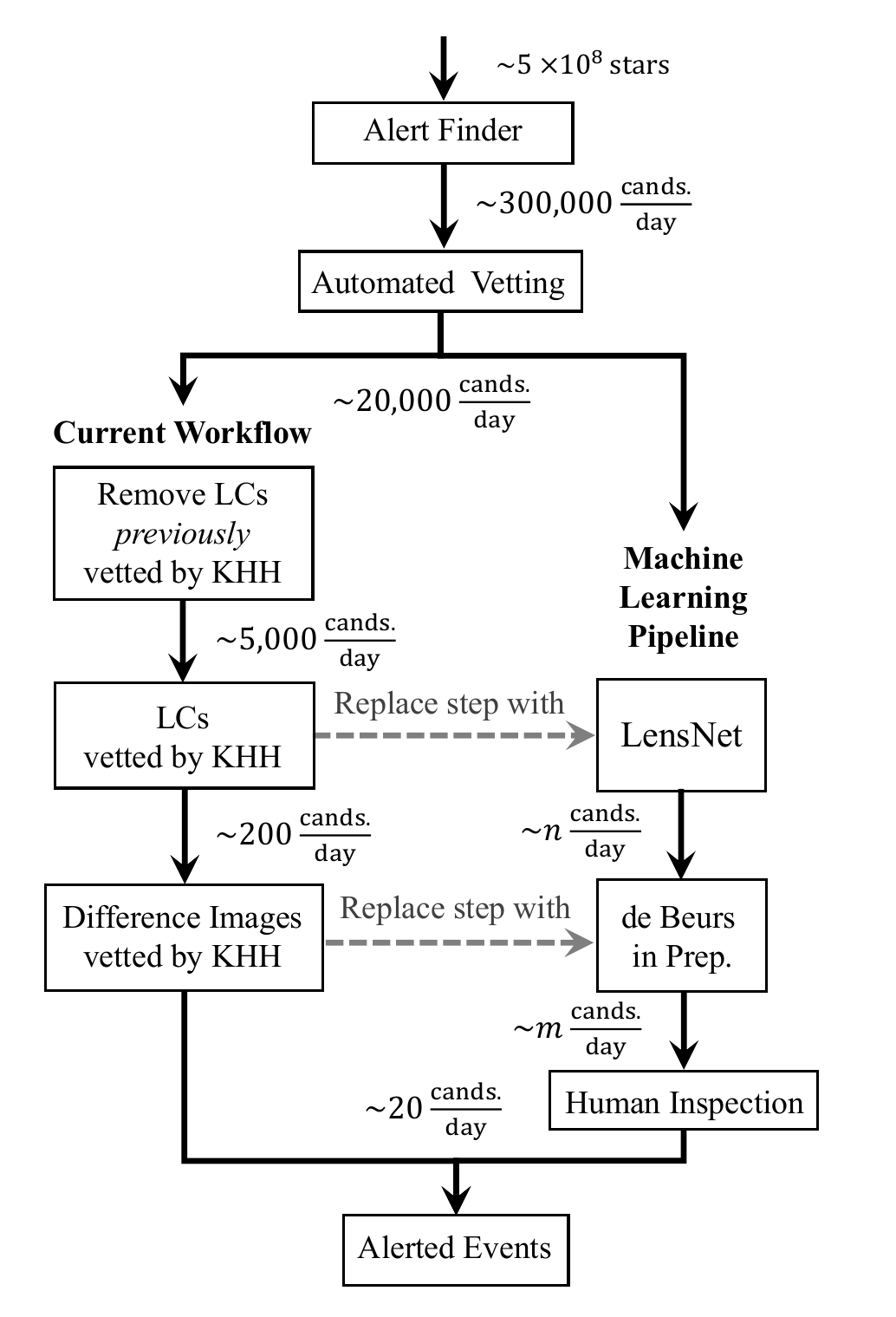}
    \caption{Comparison of the current and future pipelines for the KMTNet. The current pipeline, which processes around 5,000 alerts per day, involves manual vetting of light curves and difference images by co-author KHH. This process reduces the number of alerts to around 20 per day. The future pipeline, incorporating LensNet and additional work from \citet{deBeurs_inprep}, will process a higher volume of alerts—around 20,000 per day—while reducing the number of manually reviewed alerts.}
    \label{fig:fig_3_box_diag}
\end{figure}

The process of event identification starting from hundreds of thousands of candidates from the AlertFinder algorithm results through to the final selection of a few dozen alerts is illustrated in Fig. \ref{fig:fig_3_box_diag} and described in detail in Section \ref{sec:AF}. As can be seen in the left branch of this figure, this process relies heavily on human reviews of the candidates. The human review consists of two  stages:  first, light curves of the candidates are reviewed, and then, difference image stamps are extracted for promising candidates and examined to confirm that a given candidate is not due to an image-level artifact. Co-author KHH has primarily been responsible for developing the algorithms that handle the automated vetting, developing the procedures and algorithms for the human review, and carrying out those reviews.

In this paper, we present LensNet, a new ML architecture designed to provide fast and reliable microlensing classification tailored for the candidates identified by the AlertFinder algorithm of KMTNet. Our algorithm processes the light curves and auxiliary data of candidate microlensing events identified by the AlertFinder algorithm, which focuses on sequences with a characteristic rise in brightness. It then provides a probability estimate for each alert, indicating the likelihood that it is a genuine microlensing event. 

This work differs significantly from previous efforts in microlensing event classification for several key reasons. First, our model focuses on detecting real-time alerts rather than post-season events, a relatively unexplored area; only one other study has attempted this \citep{Godines19}. Second, while most existing methods rely on predefined statistical metrics or manually engineered features extracted from light curves, our approach processes the entire light curve directly, capturing the full temporal evolution of the data. In these previous approaches \citep{Wyrzykowski15_OGLEIII, Godines19, Mroz20_NN, wyrzykowski2016}, where static metrics such as event amplitude, timescale, and baseline variability are used, simpler models have sufficed to process the feature set (e.g., Random Forests). However, because our method leverages the full time-series data, along with contextual features, we utilize a more advanced branched deep Recurrent Neural Network (RNN) architecture, which is designed to process multiple streams of data from different telescopes simultaneously.

The structure of this paper is as follows: We give an overview of our goals in Section \ref{sec:overview}.
Section \ref{sec:AF}, gives an overview of alert-finding in the KMTNet datastream including the algorithm (Section \ref{sec:alert_algo}) and the manual vetting process (Section \ref{sec:alert_vetting}) used for identifying microlensing events. Then detail the dataset used for training the model, starting with labeling of candidates in Section \ref{sec:data} and describing data augmentation techniques and treatment of the light curves in Section \ref{sec:lc_data}. Section \ref{sec:algo} explains the LensNet architecture, a branched Recurrent Neural Network (RNN) that independently processes data from different telescopes to classify microlensing events. In Section \ref{sec:train}, we describe the training process, including hyperparameter tuning and distributed learning techniques. Section \ref{sec:results} presents the results, analyzing the model's accuracy across binary and multi-class classification tasks, while Section \ref{sec:discuss} evaluates the performance and discusses the robustness of the model and potential avenues for future work, including the integration of difference images to further enhance classification accuracy. Finally, Section \ref{sec:concl} concludes with a summary of our findings.

\section{Overview of LensNet Properties and Goals} \label{sec:overview}

LensNet is a branched architecture of three Recurrent Neural Networks that incorporate time-series inputs from all KMTNet telescopes. Each telescope's input is a sequence of vectors, where every vector represents an observation at a specific time, composed of the following features: observation time, flux, sky background, star’s FWHM, flux uncertainty, air mass, and PSF quality. LensNet processes these time-ordered vectors chronologically, leveraging its recurrent nature to capture temporal dependencies and patterns across the full sequence.

The fundamental goal of this work is to replace or significantly reduce the human effort involved in the process of vetting AlertFinder candidates and minimize the reliance on unique human expertise, while maintaining a high detection rate of alerts amenable to followup observations. The right branch illustrated in Fig. \ref{fig:fig_3_box_diag} shows how LensNet could fit into a future workflow. In particular, the goal of this paper is to try to reproduce the human light curve review. If perfectly successful, LensNet would identify the same $\sim 200$ candidates as KHH. Then, those candidates could be passed on to the second stage of the review involving inspection of difference images. This could either be handled by a human or by an additional ML component \citep{deBeurs_inprep}.

Ideally, all of the human elements of the AlertFinder workflow would be replaced by ML models that would produce the same final set of alerts that KHH would produce. In this model, as long as the $m$ candidates produced by the second ML review were the same as those produced by the human review, the value of $n$ is irrelevant so long as it is tractable to produce difference images for all of those candidates. Even imperfect performance could still be acceptable for practical implementation. For example, we could also tolerate a larger value of $m$, so long as it contained all the real microlensing events, and still achieve a substantial improvement in the number and quality of candidates requiring human review. 

The latest optimized models, trained with an augmented dataset of real observations, achieve accuracies of 87.5\% in unseen data. Ultimately, this new tool for real-time microlensing detection could help in a variety of science cases, ranging from exoplanet surveys to studies of galactic structures and black holes.

{\section{Alert Finding in KMTNet Data}\label{sec:AF}}

\subsection{KMTNet Data}

\begin{figure}[h!]
    \centering
    \includegraphics[width=\columnwidth]{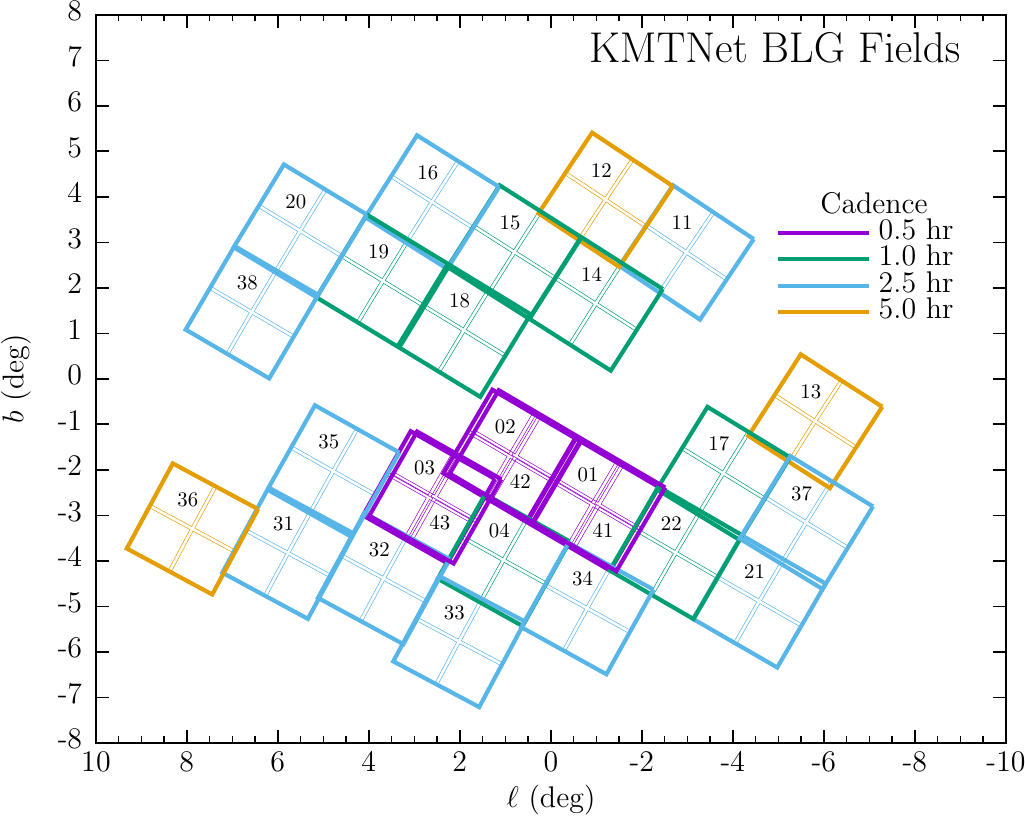}
    \caption{Map of the KMTNet BLG observing fields, showing the distribution of the four CCDs per field, with each field color-coded according to its observational cadence, ranging from 0.5 hours (purple) to 5.0 hours (yellow). The fields are plotted in galactic coordinates, longitude ($\ell$) and latitude ($b$). Credit: Matthew Penny.}
    \label{fig:fig_2_BLG_fields}
\end{figure}

As discussed in the Introduction, KMTNet combines data from three different observatories to search for candidate microlensing events. Figure \ref{fig:fig_2_BLG_fields} illustrates the tiered observing cadence strategy. KMTNet primarily observes through an $I$-band filter, although some observations are taken occasionally with a $V$-band filter. The nominal ratio of $V$ to $I$ band observations is 1:9, but in practice, there are minor variations to take advantage of a particular site's unique characteristics. The $V$-band observations are excluded from  evaluation for alerts, but they create time gaps in the $I$-band datasets. 

The imaging data is initially reduced by a difference imaging pipeline using the DIA algorithm by \citet{Wozniak00}. This produces a light curve consisting of Heliocentric Julian Dates (HJD), flux measurements, uncertainties in the flux  measurements, and a number of other diagnostic parameters (sky background, seeing, PSF $\chi^2$).

\subsection{AlertFinder Algorithm} \label{sec:alert_algo}

Here, we briefly review the main components of the KMTNet AlertFinder algorithm as it was described by \citet{Kim18_AF}.

First, the diagnostic information is used to mask the data and remove data points likely to be photometric outliers. After that, the data are reduced to just epoch, difference flux, and flux error and converted to binary format. This reduction and change in format is a practical adaptation needed to reduce memory requirements and processing time for the 500 million KMTNet light curves. These binary files are passed to the AlertFinder algorithm without the diagnostic information.

\begin{figure}[h!]
    \centering
    \includegraphics[width=\columnwidth]{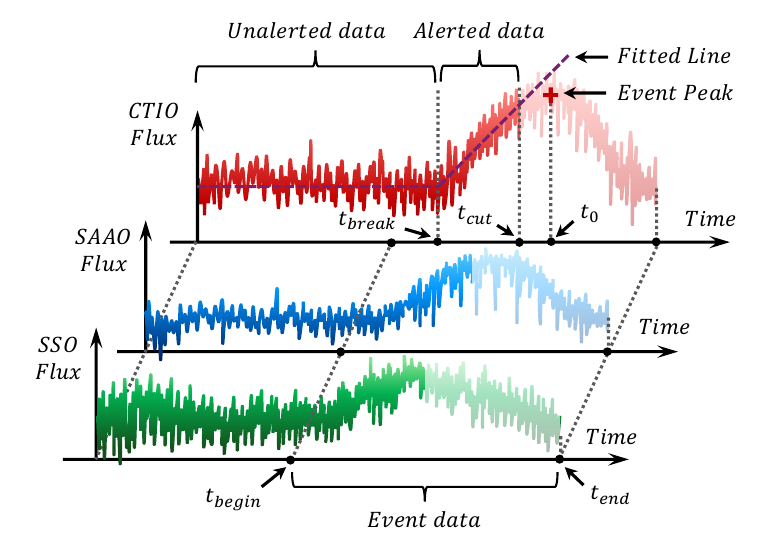}
    \caption{Demonstration of the process to identify potential microlensing events by analyzing the light curves from the three different telescopes: CTIO, SAAO, and SSO. The point $t_{\text{break}}$ represents the time at which the algorithm identifies the start of the rising event, while $t_{\text{cut}}$ is the moment at which the data was evaluated. The point $t_0$ corresponds to the peak of the microlensing event assuming it is a genuine event. In this particular case it was the CTIO telescope that triggered the alert, and thus the specific time points $t_{\text{break}}$, $t_{\text{cut}}$, and $t_0$ refer to the CTIO's time series data. The ``unalerted" data represents the baseline flux before any indication of a microlensing event, while the ``alerted" data shows a noticeable increase in flux. The fitted line over the alerted data highlights the trend in flux increase, which is used to detect the alert. The figure captures the progression of the event from its real beginning ($t_{\text{begin}}$) to its end ($t_{\text{end}}$). This figure also shows how the microlensing pattern can be ``seen" in all the three telescopes, despite the fact that is only alerted from the data of one of them.}
    \label{fig:fig_4_three_tels}
\end{figure}

The AlertFinder algorithm then evaluates all unique permutations and combinations of the datasets to check whether at least $N_{\rm high}$ of the last $(N_{\rm high} + 10)$ flux measurements are $\ge 3\sigma$ above the median flux. Finally, if this is true, the light curve is fit with a function consisting of a flat line plus a line with a constant slope (as shown in Fig. \ref{fig:fig_4_three_tels}), i.e.,
\begin{equation}
    F(t) = a_0 + a_1(t-t_{\rm break})\Theta(t-t_{\rm break})
    \label{eqn:break}
\end{equation}
where $\Theta$ is a Heaviside step function and $t_{\rm break}$ is the time of the break point. If Equation (\mbox{\ref{eqn:break}}) is a significantly better fit to the data than a flat line ($\Delta\chi^2 > \Delta\chi^2_{\rm thresh}$), then the light curve is flagged as a candidate microlensing event. Typically, each day $\sim3\times10^5$ candidate events (out of $\sim5 \times10^8$ total stellar light curves) are flagged as possible candidates.

{\subsection{Automated Vetting}\label{sec:auto_vetting}}

Many of the AlertFinder candidates are spatially correlated, because when using the DIA photometric package, a varying star will often affect light curves of nearby stars, e.g. creating ``ghost" events that have the same properties \citep{Wyrzykowski15_OGLEIII}. This effect is particularly bad for bright, variable stars, sometimes creating large clusters of false signals. Hence, any candidates with many neighboring candidates are rejected. A maximum of 20,000 candidates per field or 800 candidates per chip (1/4 of a field) are kept (sorted by star ID number, which gives preference to candidates detected in CTIO data (whose star IDs begin with ``BLG"). This process brings the total number of candidates to be reviewed down to $\sim3\times10^4$. 

These remaining candidates are divided into two categories. About 3/4 were previously reviewed by KHH in the past 6 days and not selected as microlensing events, leaving $\sim6 \times10^3$ new light curves that require manual vetting. The other 3/4 may be reviewed or not depending on the specific load and time available on a particular day. 


\subsection{Manual Vetting} \label{sec:alert_vetting}

Because of the large fraction of false positives, the default status of each light curve is set to ``No," meaning "do not alert." Each light curve is then reviewed by KHH, who flags potential real microlensing events. For any events that are flagged at this stage (around 200), light curves of the nearest neighbors are re-evaluated to identify the strongest signal (since the change in flux of any given star is frequently correlated with its neighbors). Then, the most recent six difference image stamps are extracted and visually inspected to check for obvious artifacts (such as bleed trails from saturated stars or diffraction spikes). Finally, the remaining candidates are classified in ratings from 1 to 3 (which correspond to ``clear", ``probable", and ``possible", respectively)\footnote{There is also a ``4" category for events that were previously alerted but later recognized as false positives}. Candidates in classes 1 and 2 are alerted.

\subsection{Common False Positives}

\begin{figure}
\begin{center}
    \includegraphics[width=0.4\textwidth]{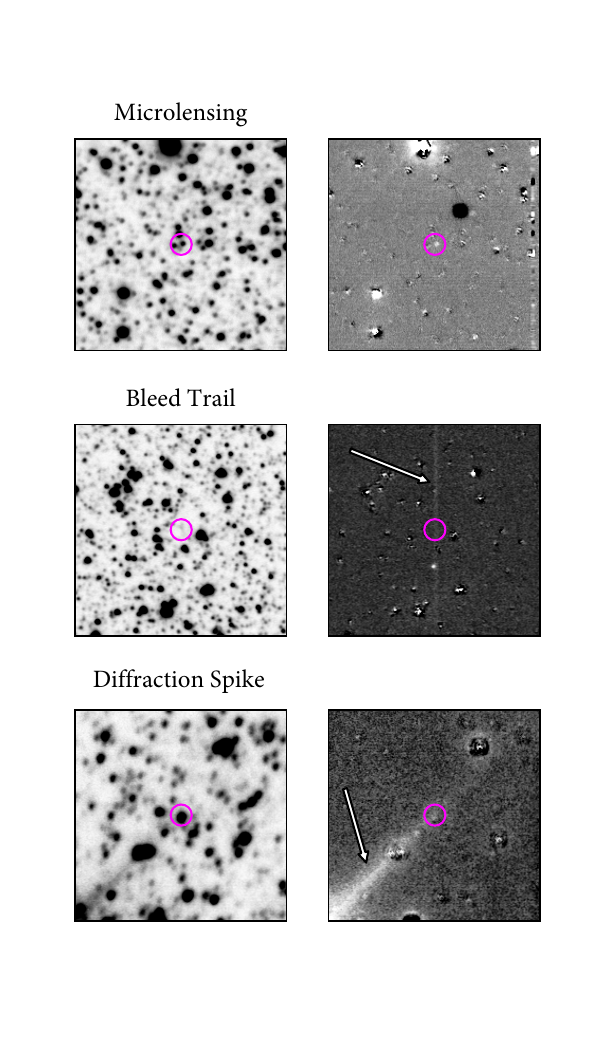}
    \caption{Examples of a real microlensing event (top) compared to two common types of false positives: a bleed trail (center) and a diffraction spike (bottom). Left panels show the original images, and the right panels show the difference images. A magenta circle indicates the location of the catalog star on which the candidate was detected. White arrows indicate the relevant feature causing the false positive.} \label{fig:fig_5_diff_ims}
\end{center}
\end{figure}

\begin{figure*}
    \centering
    \includegraphics[width=\linewidth]{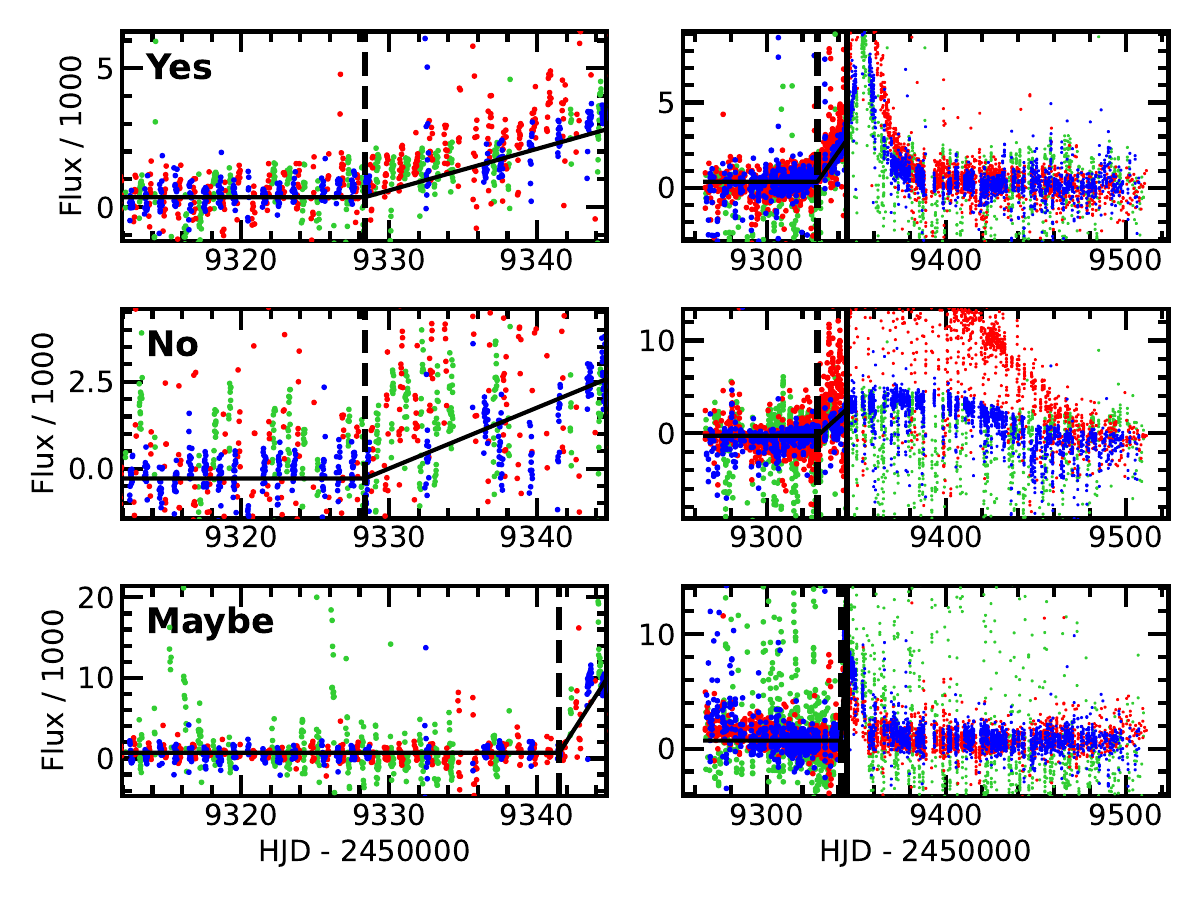}
    \caption{Light curves of a candidate found by AlertFinder on 2021 May 10 UT 06:00 (i.e., $t_{\rm cut} = 9344.75$) from each category: ``Yes" (top), ``No" (middle), ``Maybe" (bottom). Left panels show the light curves as they would appear at $t_{\rm cut}$ with the fitted broken line (black solid line). Right panels show the full 2021 light curves with smaller markers for data with $t > t_{\rm cut}$. The colors show data from different observatories: SAAO (dataset with the strongest AlertFinder signal; blue), CTIO (red), SSO (green). The vertical dashed line shows $t_{\rm break}$, and the vertical solid line shows $t_{\rm cut}$.}
    \label{fig:fig_6_lc_examples}
\end{figure*}

There are three common types of false positives for candidate events: 
\begin{itemize}
    \item  \textbf{Bleeding due to saturated stars:} 
    If a star is very bright, the number of photons received by a given pixel in the detector can exceed its capacity, and the extra photons can ``bleed" into neighboring pixels in the same column. In the images, this effect appears as vertical stripes.
    
    \item  \textbf{Diffraction spikes:} Diffraction of light inside the telescope system can result in a stellated pattern extending out from the location of a bright star.
    
    \item  \textbf{Nearby variable stars:}
    Variable stars are often very bright, so if bleeding or diffraction spikes are associated with one of those stars, then those signals will be time variable. Variable stars can also corrupt the kernel calculated for the difference imaging, so the photometry of a nearby star can be correlated with the variable even without these explicit effects.
\end{itemize}
These phenomena can affect the light curves of nearby stars, depending on their (X, Y) location within the detector relative to the star creating the effect. Because they depend on many factors, the effects these phenomena produce can be time-variable, creating temporary increases in brightness in the light curves of nearby stars that are detected as candidate microlensing events.
A comparison of a real microlensing event, a false positive due to a bleeding trail, and a false positive due to a diffraction spike can be seen in Fig. \ref{fig:fig_5_diff_ims}. The middle set of light curves in Fig. \ref{fig:fig_6_lc_examples} show an example of contamination by a nearby, bright variable star.

Many of these common false positives can be eliminated during the manual vetting of light curves, and further refinement is typically achieved by analyzing difference images. However, true microlensing events can still occur in proximity to these effects, making the identification process more challenging.

\section{Label Generation \& Training Set Assembly} \label{sec:data}

For our training sample, we use data from the 2021 KMTNet microlensing season. For each date the AlertFinder was run (weekdays from approximately April--September),  there are lists of the $\sim300,000$ stars selected by the AlertFinder algorithm on that date as possible microlensing events, $\sim 20,000$  stars selected for light curve review, $\sim 200$  candidates selected for difference imaging review, and $\sim 20$ actual alerts from that date.

Since the LensNet pipeline will be used on the sample of $\sim20,000$ candidates selected for light curve review (Fig. \ref{fig:fig_3_box_diag}), we selected our training sample from that group. In the human review, as discussed in Section \ref{sec:auto_vetting}, usually only 1/4 of these candidates are reviewed because the other 3/4 had been previously reviewed and not selected by a human, but this is not a metric we can apply to a purely ML algorithm. 

There are several challenges in constructing a training set for the ML algorithm. First, the distribution of labels in the underlying database is severely biased toward false positives. Second, recreating the DIA  light curves is computationally intensive. Given these constraints, we needed to curate these lists of AlertFinder candidates into a more-balanced, tractable training set.

Based on the 2-stage manual review, we define three possible classifications for a given candidate:
\begin{itemize}
  \item{``No'': candidate was rejected as false during the light curve review,}
  \item{``Maybe'': candidate was identified as a possible event during light curve review but deemed to be a false positive after examining the images,}
  \item{``Yes'': candidate passed both the light curve and image review and was announced as an event.}
\end{itemize}
An example of a light curve in each category is shown in Fig. \ref{fig:fig_6_lc_examples}.
Given our immediate goal to try to reproduce the results of the human light curve review, candidates labeled ``Maybe" could be grouped with the ``Yes" candidates. In our study, we evaluate both the 3-class labeling task (``No" vs. ``Maybe" vs. ``Yes") and 2-class labeling task (``No" vs. ``Maybe" + ``Yes").

To construct our training set, we identified a set of candidates in each category and then extracted their DIA light curves for training the RNN. We used candidate lists from 10 April 2021 to 26 Sept 2021. KMTNet started alerts earlier than 10 April, but this limit ensures that there are a reasonable number of data points in a given light curve for training the RNN.

\subsection{``Yes''}

Initially, we created the ``Yes" list by selecting any candidates that were
flagged as [1, 2, 3, 4] (i.e., [`clear', `probable', `possible', `not-ulens']) and cross-referencing them with the KMTNet event table. This resulted in 4183 candidates, but since some of them were flagged on multiple dates before being alerted, there are only 2158 unique stars in this list. For duplicate stars, we used the last review date ($\max(t_{\rm cut})$).

Then, we removed any events that were not classified as either ``clear'' or ``probable'' microlensing events in both the EventFinder \citep{KimKim18_EF, Kim18EF} and AlertFinder searches. For reference, the KMTNet EventFinder algorithm differs from the AlertFinder algorithm in that it is run after the microlensing season concludes (so on the full season's data) and, because of that, it uses a full Paczy\'{n}ski microlensing light curve model for the fitting rather than a broken line (see \citet{KimKim18_EF, Kim18EF} for more details).

In this sample, we also checked for duplicate events (alerted based on different catalog stars, but at the same coordinates), but did not find any.

\subsection{``Maybe''}

There are multiple reasons a candidate might be flagged as a possible candidate during the light curve review stage but not alerted:
\begin{enumerate}
  \item{It is a real event but it was not clear until more data were taken,}
  \item{It is a ``ghost'' event: it is near a real event, so the light curve shows the ``echo" of a real event due to contamination,}
  \item{It is a true false positive identified during difference image review, e.g., due to a bleed trail.} 
\end{enumerate}
Because candidates in the first two categories share many light curve characteristics with real microlensing, we eliminated them from the ``Maybe'' sample. For category (1), this is straight forward and just involves checking for a star in the KMTNet event table with the same ID. For category (2), we cross-checked the coordinates of the candidate against all events in the KMTNet event table and eliminated the candidate if it was within $20^{\prime\prime}$ of an object in the table.

Then, as with the ``Yes'' category, because a given ``Maybe'' candidate might be appear on multiple dates, we only kept unique stars and used the last review date ($\max(t_{\rm cut})$). This leaves 1361 candidates in the ``Maybe'' sample.

Note for this sample, we did {\it not} remove any candidates due to proximity to other candidates in the ``Maybe'' sample. As a result, some candidates in the ``Maybe'' sample are spatially correlated. This reflects a real effect, e.g., bleed trails affect stars in the same column in a similar way. Hence, these correlations reflect real features in the data that can be used to identify objects in this category, even though we do not include spatial position as a training feature.

\subsection{``No''}

There are an abundance of candidates in the ``No'' category. Thus, we chose a random sample of a similar size of the ``Yes'' and ``Maybe'' samples.

We start by choosing ``No'' candidates that were alerted on the same date as the initial ``Yes'' sample. From the ``No'' candidates, we randomly select a number of ``No'' candidates equal to the total number of ``Yes'' candidates from that date. Candidates within $20^{\prime\prime}$ of a ``Yes'' candidate from the same date are excluded from this selection. From this random selection, we randomly down-selected to 2098 ``No'' candidates (this particular number was chosen to match the number of unique ``Yes'' candidates prior to filtering for event quality) and eliminated any candidates within $20^{\prime\prime}$ of a real event. This leaves 2065 candidates in our ``No'' sample.

While we do check for duplicated stars in this sample, we do not check for proximity to other candidates in the ``No'' sample.

\section{Light Curve Preparation for ML Pipeline} \label{sec:lc_data}

After assembling the labeled dataset, several preprocessing techniques are applied to clean, augment, and standardize the data. The time series data, which form the basis of the microlensing event classification task, undergo multiple steps, including handling missing values, normalizing features, and ensuring proper alignment across different telescopes. In addition, we use a data augmentation technique to artificially expand the training set by cropping the time series on either side of the microlensing event, enhancing the model’s generalization ability. This ultimately helps simulate various observational conditions, allowing the model to perform robustly even with incomplete or noisy data. The remaining preprocessing steps ensure that the data is consistently prepared for model training.

\subsection{KMTNet Data Properties: Implications for Use as Neural Network Inputs}

The observations from different sites are reduced and analyzed separately for every field, due to offsets, varying systematics, and other site-specific discrepancies. As a result, most stars have three associated data files, while stars in overlapping fields may have six or more. Furthermore, every site implements slight deviations from the nominal observing strategy to leverage its unique characteristics. Thus, the specific properties of each time series data file—such as its length, spacing, and quality—depend on several factors. For example, weather conditions at a particular site may prevent observations or degrade their quality. Unlike many ML problems that work with uniformly sampled high-quality data, KMTNet data contains irregularities in both the timing and quality of our observations. Consequently, preparing the data for use in an ML algorithm requires a more sophisticated approach to data processing in order to accommodate these inconsistencies.

Therefore, we decided to separate the different telescope data inputs when processing them through our model. We designed LensNet as a branched pipeline that analyzes each data file individually, accounting for variations in observation conditions like systematics and site-specific characteristics. This approach allows the model to capture subtle distinctions between datasets that a single unified process would overlook, ultimately making a better use of the unique information each observing site provides, and thus leading to more accurate and reliable predictions.

\subsection{Data Cleaning}

We perform a thorough cleaning and curation process of our dataset. First, we impose a limit on the sequence length, where each candidate must have no more than 1,500 observations. Sequences longer than this length are cropped, because in the majority of cases, the model is already able to capture the patterns of the star within this length. Furthermore, the additional predictive power from the remaining data is relatively limited and comes at a high computational cost. 
Moreover, we handle duplicate instances carefully: if a star appears in two or more categories in different days, it is reassigned to a single category (we perform this reassignment is handled on a case-by-case basis).

In addition, we apply specific constraints to the data: We remove observations where the flux error is negative, where the PSF quality is outside the acceptable range of 0 to 100, and where the FWHM is negative. We also require that a candidate star has a minimum of 10 valid observations from at least one of the telescopes after all the cleaning steps.

Following this process, the curated dataset consists of 1,190 instances in the ``Maybe" category, 1,825 instances in the ``Yes" category, and 2,038 instances in the ``No" category. While these classes are not perfectly balanced, we account for this imbalance by applying different weights to adjust the learning rates for each category during model training, ensuring that the classifier learns effectively across all classes.

\subsection{Data augmentation} \label{sec:augm}

Data augmentation is a popular technique used to artificially expand the size and diversity of a dataset by applying various transformations to the original data. This process helps improve the model's generalization ability by exposing it to a wider range of possible input scenarios. In time-series data, augmentation can involve modifying the sequences in ways that mimic real-world variations, enhancing the model's ability to handle different observational conditions.

\begin{figure}[h!]
    \centering
    \includegraphics[width=\columnwidth]{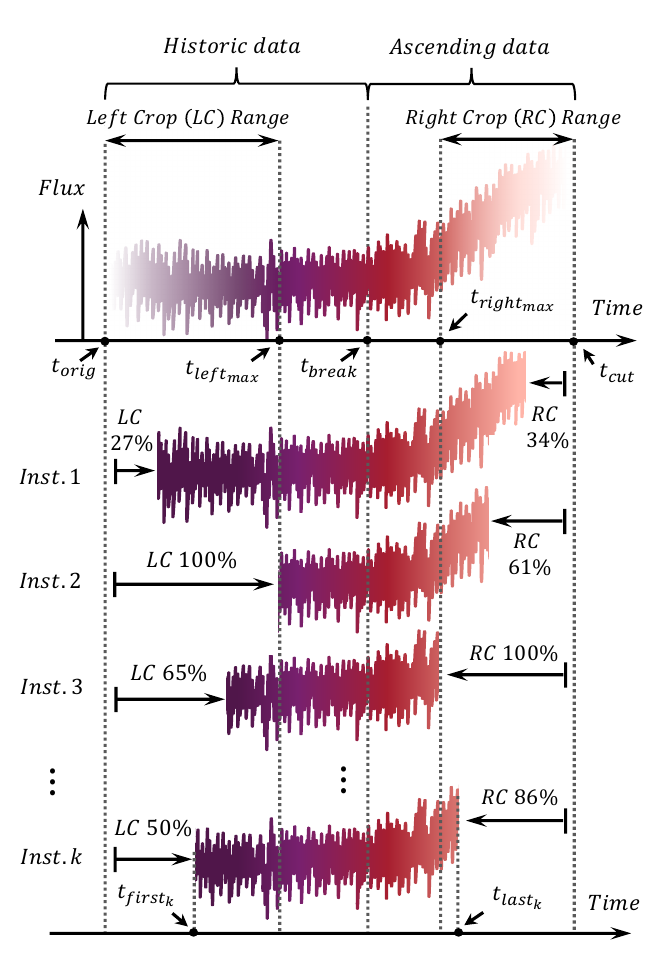}
    \caption{Illustration of the data augmentation process applied to the flux time series and associated features for the training of the microlensing event classification model. The figure shows multiple augmented flux instances of the same real data from a given instrument, where the time series has been cropped from the left (historic data) and right (ascending flux data) sides. The percentages indicate the extent of the data removed in the cropping of each side. This augmentation simulates varying levels of prior knowledge (left cropping) and available microlensing alert information (right cropping), ensuring the model is robust to different observational scenarios.}
    \label{fig:fig_7_data_aug}
\end{figure}

To enhance the robustness of our microlensing event classification model, we applied a data augmentation technique by cropping the time series data on both the left (historic data) and right (ascending flux data) sides (see Fig. \ref{fig:fig_7_data_aug}). The left cropping reduces the amount of prior knowledge about the behavior of the background star before the suspected microlensing event, while the right cropping limits the available data during the event's rise in flux. This approach is particularly valuable because alerts for microlensing events may occur with varying amounts of historic or ascending flux data, reflecting real conditions where such variations are common.

In addition to cropping the flux time series, we also applied the same cropping to the other features (flux error, air mass, FWHM, etc.) ensuring consistency across all inputs to the model. Importantly, a minimum data retention threshold was enforced on both sides to maintain sufficient information for model training. Logically, this augmentation process was applied uniformly across data from all three telescopes, so that the cropping intervals were equivalent, thus preserving the temporal alignment of the observations.

This method not only broadens the diversity of the training dataset but also aligns with the nature of recurrent neural network architectures, which benefit from learning temporal dependencies across varying data lengths. By training with these augmented datasets, we obtain a better equipped model capable of handle real-time alerts with different amounts of available data, thus enhancing its overall performance. Table \ref{tab:table_1} summarizes the number of instances before and after the augmentation process.

\begin{table}[h!]
    \centering
    \begin{tabular}{lrr}
        \toprule
        \textbf{Category} & \textbf{Non-Augmented} & \textbf{Augmented} \\ 
        \midrule
        Maybe    & 1,190  & 7,404  \\
        No       & 2,038  & 12,008 \\
        Yes      & 1,825  & 10,902 \\ 
        \midrule
        \textbf{Total}  & 5,053  & 30,314 \\
        \bottomrule
    \end{tabular}
    \caption{Comparison of non-augmented and augmented instances per category, including totals. Imbalance difference after the augmentation was intentional to increase the representation of the ``Maybe" category in the dataset, as it is more difficult to classify than the other categories.}
    \label{tab:table_1}
\end{table}

Table \ref{tab:table_2} provides a comparison of the number of instances and the fractions of examples across the training, validation, and testing datasets, both with and without augmentation. It is important to note that while the data are split based on individual stars, the augmentation process generates a variable number of instances per star, resulting in varied instance counts across the different datasets.

\begin{table}[h!]
    \centering
    \begin{tabular}{lrr}
        \toprule
        \textbf{Dataset} & \textbf{Instances} & \textbf{Real Fraction} \\ 
        \midrule
        Training (Aug)    & 24,276 & 0.80082  \\
        Validation (Aug)  & 3,014  & 0.09943  \\
        Testing (Aug)     & 3,024  & 0.09976  \\
        \midrule
        \textbf{Total (Aug)}     & 30,314  & 1.000  \\
        \midrule
        Training (Non-Aug)   & 4,046  & 0.80071  \\
        Validation (Non-Aug) & 503    & 0.09954  \\
        Testing (Non-Aug)    & 504    & 0.09974  \\
        \midrule
        \textbf{Total (Non-Aug)} & 5,053  & 1.000  \\
        \bottomrule
    \end{tabular}
    \caption{Comparison of instances and real fractions for augmented and non-augmented datasets, including totals.}
    \label{tab:table_2}
\end{table}

The data augmentation process was applied only after splitting the dataset into training, validation, and test sets to ensure no data leakage between them. This step was crucial to avoid artificially inflating the model's performance and to ensure that the model's ability to generalize was tested on truly unseen data.

Along these lines of ensuring fairness in the evaluation, we also ensured that no data from the same star was present in multiple sets simultaneously. By carefully splitting the dataset this way, we guaranteed that each star's data was unique to a single set, preventing the model from learning specific patterns related to the same star across different sets. This approach ensures a more rigorous evaluation of the model's ability to generalize to entirely new microlensing alerts.

\subsection{Data Pre-processing Pipeline} \label{sec:prepro}

The preprocessing pipeline for the time series data consists of several crucial steps that ensure the input data is standardized, free from outliers, and properly aligned for the Recurrent Neural Network (RNN) architecture. This section details the steps of time relativization, outlier removal, NaN handling, fitting, standardization, and padding, each of which contributes to the robustness and accuracy of the model.

\begin{itemize}
     
    \item \textbf{Time Relativization:} To standardize the temporal dimension of the data across different telescopes, we implement a time relativization step. Here, the time values in the dataset are adjusted relative to the moment of the last observation across all three telescopes, denoted as $t_{\text{last}}$. This point is crucial as it represents the time at which the alert was triggered. However, this relativization is performed after data augmentation to avoid any bias or leakage of future information into the model. By shifting the time axis in this manner, we ensure that the model's input data is aligned temporally, regardless of the specific observational timelines of each telescope.
    
    \item \textbf{Outlier Removal:} After time relativization, we apply an outlier removal algorithm to clean the data. This step involves identifying and excluding data points that deviate significantly from the overall trend, both upwards and downwards. The removal is conducted across all features. This process helps eliminate anomalies that could negatively impact the model’s performance, ensuring that the input data more accurately represents the underlying astrophysical signals.
    
    \item \textbf{NaN Handling and Filling:} We remove any missing data points (NaNs) that are present in any of the features during preprocessing, rather than attempting to fill gaps through interpolation. Given that the Recurrent Neural Network (RNN) architecture processes data sequentially, interpolating missing values would not provide any additional information or benefit to the model. Moreover, because the RNN is not constrained by a fixed input length, we are able to preserve the integrity of the remaining data without the need for imputation. This flexibility is one of the key advantages of the chosen RNN architecture, allowing us to work with variable-length sequences while ensuring the model is not disrupted by incomplete observations or gaps.
    
    \item \textbf{Fitting the Ascending Data:} The next step in the preprocessing pipeline involves fitting a line to the ascending data portion of the flux, which is critical for capturing the microlensing event’s rising phase (this step implicitly mimics the AlertFinder algorithm behavior). The fitting process takes into account the flux error, providing a weighted fit that more accurately reflects the observational uncertainties. The fitted line is then treated as an additional feature in the input vector that is passed to the ML algorithm. Specifically, during the historic (pre-event) data, this feature is set to the average flux in that region. After $t_{\text{break}}$, the feature transitions to match the values of the fitted line, creating a piecewise smooth function that serves as a robust input for the model. A key constraint imposed during fitting is that the line must start at the point of $t_{\text{break}}$ and the average historic flux, ensuring consistency and continuity in the data representation.
    
    \item \textbf{Standardization:} To ensure that the data from all telescopes is on a comparable scale, we apply standardization independently to each feature. This involves normalizing the data by subtracting the mean and dividing by the standard deviation, calculated from the training set. These normalization parameters are then applied uniformly across the training, validation, and test sets. For the piecewise fitted line, standardization is performed using the mean and standard deviation of the flux data to prevent decoupling between the fitted line and the actual flux values. Standardization is critical for the convergence and stability of the RNN during training.
    
    \item \textbf{Padding:} Finally, we apply padding to the time series data, a necessary step in preparing the data for input into the RNN. In our pipeline, padding involves appending zeros to the beginning of each sequence so that all input sequences have the same length, regardless of the actual duration of observations. This ensures that the RNN processes uniform-length sequences, which is essential for batch processing and model efficiency. The padding is carefully managed to avoid introducing artifacts into the data, with the padded sequences still maintaining the integrity of the temporal relationships within each telescope's dataset.
    
\end{itemize}

These preprocessing steps are essential for transforming the raw observational data into a format that is suitable for input into the RNN architecture, ultimately enhancing the model’s ability to accurately detect and classify microlensing events.

\section{LensNet Architecture} \label{sec:algo}

\begin{figure*}[h]
    \centering
    \includegraphics[width=\textwidth]{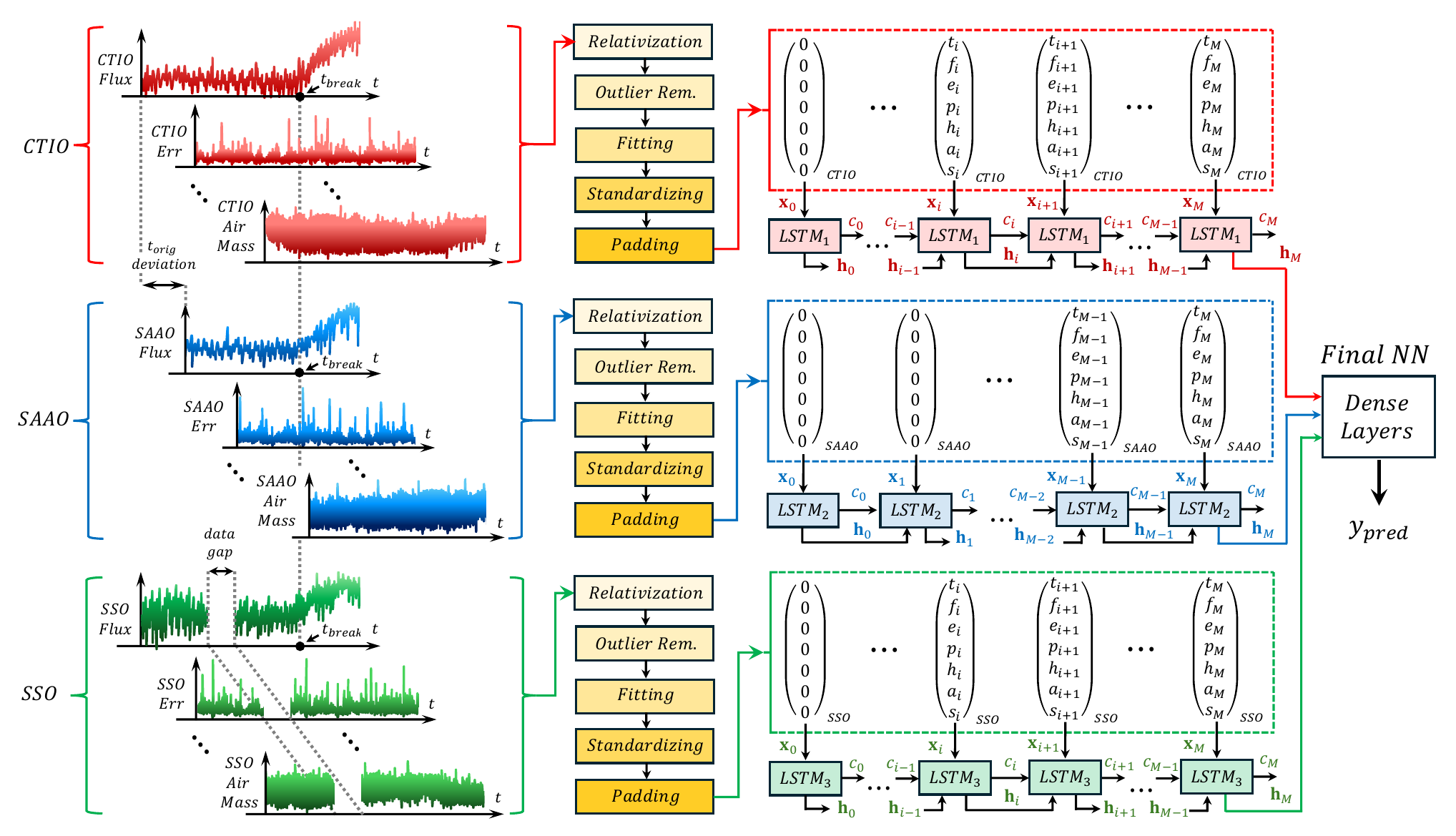}
    \caption{Schematic of the branched Recurrent Neural Network architecture used to classify microlensing alerts based on time series data from three telescopes: CTIO, SAAO, and SSO. Each telescope's data is processed independently through a dedicated RNN branch that handles a vector of multiple features (relative time, flux, flux error, air mass, FWHM, PSF, and piecewise fitted flux). The figure also illustrates the key pre-processing steps, including time relativization, outlier removal, feature fitting, standardization, and padding, which prepare the data for input into the RNN. Additionally, the unrolled RNN is depicted, showing how hidden states are passed from one time step to the next, capturing the temporal dependencies within the data. The figure highlights differences in observational data, such as the later start of SAAO observations and data gaps in SSO, which justify the use of separate RNN branches. The encoded outputs from each RNN branch are then combined and passed through dense layers to predict whether the event is a microlensing event. This architecture is designed to handle the varying quality and temporal resolution of data across the three telescopes effectively.}
    \label{fig:fig_8_LensNet}
\end{figure*}

To accurately classify potential microlensing events using time series data from the three different telescopes (CTIO, SAAO, and SSO), we designed LensNet, a branched Recurrent Neural Network (RNN) architecture where each branch processes data from a single telescope independently. A schematic pipeline of how LensNet works is depicted in Fig. \ref{fig:fig_8_LensNet}. This design choice is motivated by the unique observational characteristics and challenges presented by each telescope's data. For instance, in Fig. \ref{fig:fig_8_LensNet}, the CTIO telescope has the most comprehensive coverage, while SAAO starts its observations later (leading to a shorter time series), and the SSO data contains a gap in the observations.

RNNs are particularly well-suited for this task because they are designed to handle sequential data, making them ideal for analyzing time series where temporal dependencies are crucial. Furthermore, by treating each telescope's data separately, we ensure that the model can adapt to the specific noise patterns, data gaps, and observational cadences of each instrument. Each branch of the RNN processes not only the flux data but also other key features which allows the network to capture the complex relationships between these variables over time. These features include the observation time, the flux error, air mass, FWHM, $\chi^2$, and the PSF quality.

After processing the data from each telescope through its respective RNN, the final encoded states are combined and passed through a series of dense layers to produce the final classification output, which then determines if the alert is a microlensing event. This approach allows the model to leverage the full temporal resolution of each telescope's observations while maintaining flexibility in dealing with the varying lengths and quality of data streams.

In addition to the branched RNN architecture, we explored several alternative methods to classify the microlensing events. These included extracting features from the fitted piece-wise linear function and utilizing a variety of classical ML algorithms, such as XGBoost, bagging, decision trees, and perceptron models. The features we considered included the average standard deviation of the flux relative to the fit before and after the microlensing event (a proxy for fitting error pre- and post- $t_{break}$), the slope of the rising flux, parameters from higher-order fittings, skewness, kurtosis, and the number of points before and after $t_{break}$. Despite these efforts, we did not observe any significant correlation between these features and the target labels, with the models performing at around $\sim50\%$ binary accuracy, essentially equivalent to random guessing.

We also experimented with using an RNN architecture that only processed CTIO data, given its higher confidence in many of the training instances, as well as employing convolutional architectures to process the time series data. Ultimately, however, the architecture we present in this paper—the branched RNN that processes data from each telescope independently—outperformed all the other approaches studied, offering the most accurate and robust results.

\section{LensNet Training} \label{sec:train}

We conducted an in-depth study on two distinct classification tasks. The first task was a binary classification problem, where we combined the ``Yes" and ``Maybe" labels into a single class, distinguishing them from the ``No" class. This approach was based on the understanding that ``Yes" and ``Maybe" are often indistinguishable in flux data alone and are typically sub-classified manually only after reviewing the difference image. This mimics the classification performed by co-author KHH. The second, more challenging task was a multi-class classification problem, where the model aimed to differentiate between ``Yes," ``Maybe," and ``No" classes independently. This task was particularly difficult because even expert human vetters often struggle to distinguish between ``Yes" and ``Maybe" based solely on flux data. However, we hypothesized—and subsequently demonstrated—that the time series feature patterns contain subtle yet valuable information that can enable successful multi-class classification, even without the need for a difference image.

The model was trained using a distributed learning approach to leverage the computational power of multiple GPUs. We used TensorFlow \citep{tensorflow2015-whitepaper} for the implementation, and training was conducted on the Massachusetts Institute of Technology's (MIT) Engaging supercomputer, a high-performance computing environment designed for large-scale, data-intensive tasks. To optimize the training process, the job was performed with 4 NVIDIA A100s GPUs accelerating the training by enabling parallel processing of large data batches, 128 CPU cores, and 100GB of RAM. The job was executed on a single node to minimize communication latency between GPUs. 

We explored various combinations of hyperparameters using Tensorboard (package inside TensorFlow). These include dropout rates, number of encoding features of the LSTM branches, number of neurons of the final dense layers, number of layers, batch sizes, learning rates, etc.

The final configuration had a computational cost of approximately 1500 training epochs per day with batches of 2048 instances.

The learning rate was set to $10^{-6}$ and we used the Adam optimizer \citep{kingma2014adam}. For the binary classification task, we employed the Binary Cross-Entropy loss function \citep{mao2023cross}, while for the multi-class classification task—predicting ``Yes," ``No," and ``Maybe"—we opted for the Categorical Cross-Entropy function \citep{mao2023cross}.
  
Class imbalance in the training data was addressed by computing class weights \citep{xu2020class}, which were used during training to ensure that the model was not biased towards the majority class.

\begin{figure}[h!]
    \centering
    \includegraphics[width=\columnwidth]{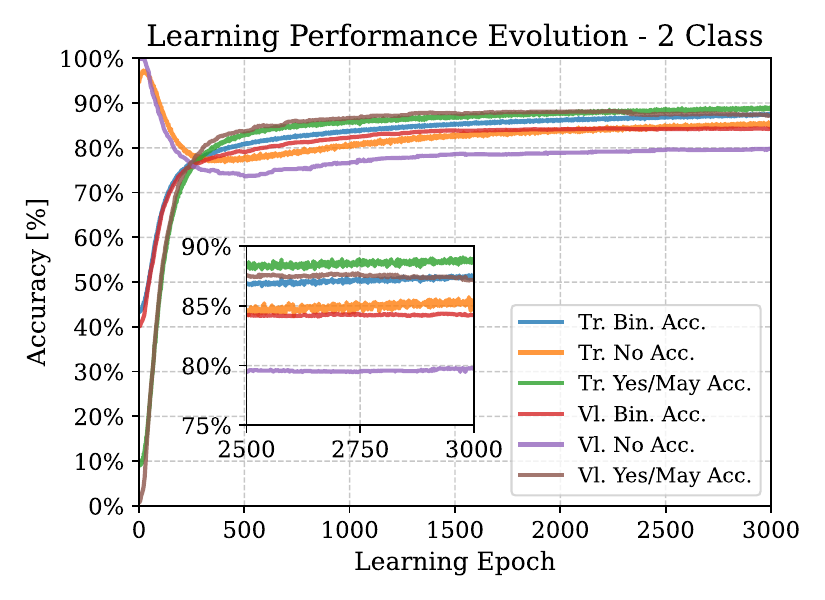}
    \caption{Learning performance evolution for the binary classification task over 3000 epochs. The graph shows the weighted, No, and ``Yes"/``Maybe" category accuracies for the training (Tr.) and validation (Vl.) sets. The model initially experiences a sharp increase in accuracy for both categories, followed by a gradual learning process that stabilizes around the 1000th epoch. The inset zooms in on the accuracy range between 75\% and 90\%, providing a closer view of the learning performance between epochs 2500 and 3000.}
    \label{fig:fig_9_acc_evol_bin}
\end{figure}

\begin{figure}[h!]
    \centering
    \includegraphics[width=\columnwidth]{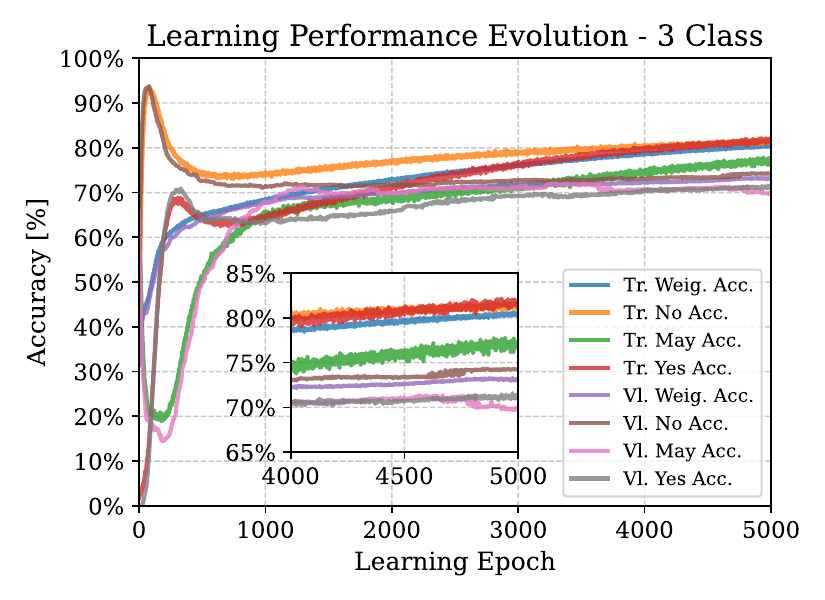}
    \caption{Learning performance evolution for the 3-class classification task over 5000 epochs. The plot illustrates the weighted, ``No", ``Maybe", and ``Yes" category accuracies for both the training (Tr.) and validation (Vl.) sets. Similar to the 2-class task, the model demonstrates an initial steep learning curve, stabilizing around the 2000th epoch. The inset highlights the performance between 65\% and 85\%, focusing on the accuracy values from epoch 4000 to 5000, where the model achieves a consistent performance across all categories.}
    \label{fig:fig_10_acc_evol_mult}
\end{figure}

Figs. \ref{fig:fig_9_acc_evol_bin} and \ref{fig:fig_10_acc_evol_mult} illustrate the learning performance evolution of the model for the binary and 3-class classification tasks, respectively. These figures highlight the progression of accuracy across categories for both the training and validation sets throughout the learning process.

In the binary classification task (Fig. \ref{fig:fig_9_acc_evol_bin}), the model shows a rapid increase in accuracy, with performance stabilizing relatively early around the 1000th epoch. The weighted accuracy remains consistently high after this point, indicating that the model quickly converges to a stable solution for distinguishing between the ``No" and ``Yes"/``Maybe" categories. This faster convergence is expected, as the binary task presents a simpler challenge compared to the 3-class task.

In contrast, the 3-class classification task (Fig. \ref{fig:fig_10_acc_evol_mult}) exhibits a more challenging learning process. While there is a sharp initial improvement in the accuracy of all categories within the first few hundred epochs, the model takes significantly longer to stabilize compared to the binary task. The weighted accuracy plateaus around 80\%, but category-specific accuracies for ``No", ``Maybe", and ``Yes" follow distinct learning trajectories. By the 4000th epoch, the model has largely stabilized, though slight improvements continue throughout the remaining epochs.

The added complexity of distinguishing between three categories, as opposed to two, is evident in the longer stabilization period. Nonetheless, the model’s performance in both tasks demonstrates its ability to adapt and achieve a high accuracy.

\section{Results} \label{sec:results}

\subsection{Performance Evaluation and Key Metrics}

\begin{figure*}[h]
    \centering
    \includegraphics[width=\textwidth]{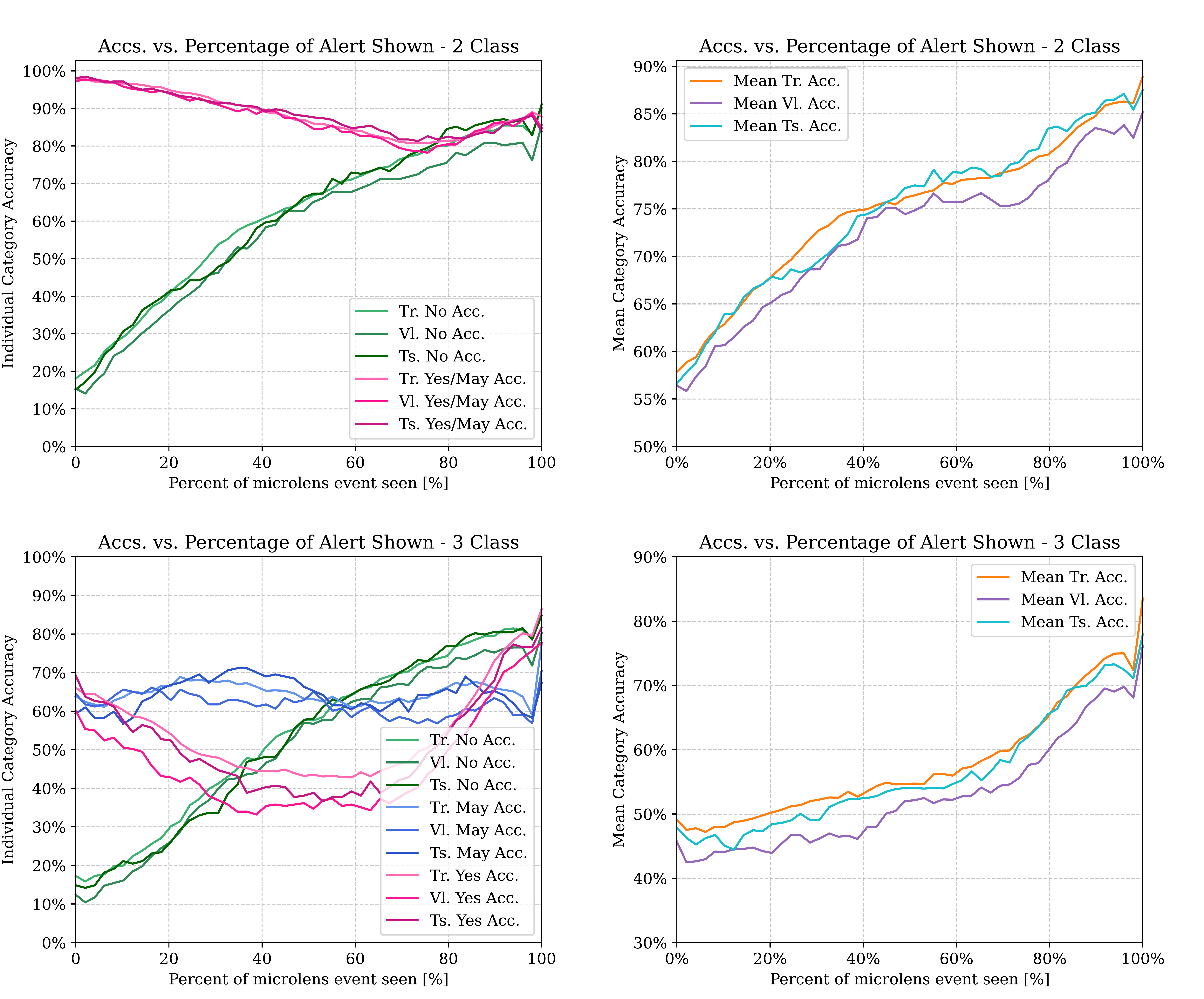}
    \caption{Evolution of categorical and mean accuracy metrics as a function of the percentage of the microlensing alert data presented to the model. The left-side panels display individual categorical accuracies for the binary (top) and multi-class (bottom) classification tasks across training (Tr.), validation (Vl.), and test (Ts.) sets. The right-side panels show the corresponding mean accuracy for each set, computed by averaging the individual categorical accuracies. The model’s performance improves steadily as it is exposed to more of the microlensing event, with sharp gains in the binary task observed between 0\% and 40\% visibility, while the multi-class task shows its most significant improvement between 60\% and 100\%. At 100\% visibility, the models achieve peak accuracies of $\sim 85\%$ for the binary task and $\sim 78\%$ for the multi-class task, demonstrating the model's ability to effectively leverage full alert visibility in real-world scenarios.}
    \label{fig:fig_11_acc_per_percentshown}
\end{figure*}

\begin{figure}[h!]
    \centering
    \includegraphics[width=\columnwidth]{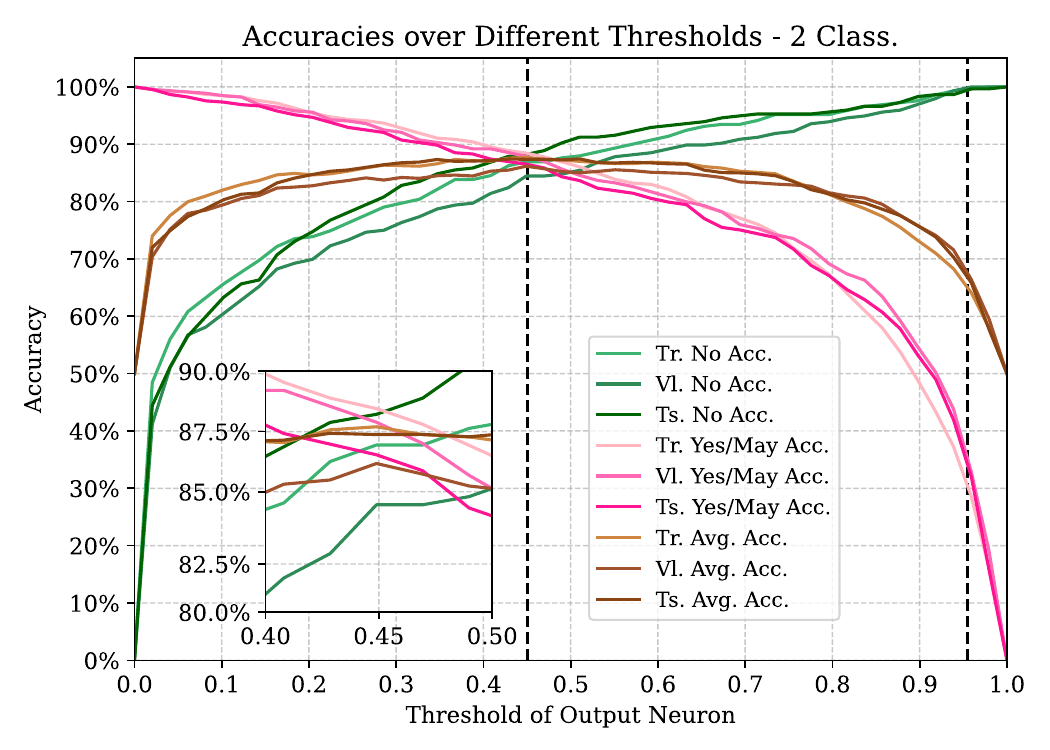}
    \caption{Accuracy over different thresholds of the output neuron for the binary classification task. The horizontal axis represents the threshold applied to the model's output neuron, while the vertical axis shows the corresponding accuracy. Green curves denote the accuracy of the ``No" category across the training, validation, and test sets, while red curves represent the accuracy for the ``Yes"/``Maybe" category. brown curves show the average accuracy across both categories, starting and ending at 50\%. The peak of the average accuracy, observed at a threshold of approximately 0.45, indicates the optimal trade-off between correctly classifying instances in both categories. However, operationally, a higher threshold is chosen to prioritize the purity of ``Yes"/``Maybe" predictions, minimizing false positives despite a higher ``No" classification rate. The vertical dashed lines mark the two thresholds utilized in the system, 0.45 for optimality and 0.955 for ``Yes"/``Maybe" purity.}
    \label{fig:fig_12_accs_per_thres}
\end{figure}

The results of the LensNet model demonstrate its strong performance in classifying microlensing events across both binary and multi-class classification tasks. In the binary task, the model effectively distinguishes between genuine potential microlensing events (grouping both ``Yes" and ``Maybe" labels) and non-events (``No"), achieving a peak accuracy of 87.5\% on the test set. In the multi-class task, which differentiates between ``Yes", ``Maybe", and ``No" labels, the model attains a lower but still robust accuracy of 78\%, reflecting the greater difficulty of separating the ambiguous ``Maybe" class from confirmed events and false positives. Throughout the experiments, the model demonstrated its robustness to partial data visibility, particularly excelling when provided with more comprehensive data (Fig. \ref{fig:fig_11_acc_per_percentshown}). Additionally, a threshold analysis of the binary task showed that higher thresholds can be used to achieve near-perfect classification purity for non-microlensing events, minimizing false positives and improving operational efficiency (Fig. \ref{fig:fig_12_accs_per_thres}). These results underscore LensNet’s potential for real-time deployment, offering a significant reduction in manual vetting while maintaining high classification accuracy.

In addition to the accuracy metrics discussed, the confusion matrices of Appendix \ref{sec:confmat} (Fig. \ref{fig:fig_13_conf_mats}) provide further insights into the model's classification performance. These matrices confirm that the model performs particularly well, especially on unseen data. We present results for both the binary classification task (with threshold values of 0.45 and 0.955) and the 3-class classification task, highlighting the model's flexibility in balancing precision and recall.

\subsection{Evaluation Under Partial Data Visibility} \label{sec:results_robust}

Fig. \ref{fig:fig_11_acc_per_percentshown} summarizes the accuracy metrics across different categories and sets as a function of the percentage of the microlensing raw non-augmented data fed into the network. This experiment allows us to evaluate the performance of LensNet with varying amounts of available information. Specifically, the horizontal axis in all the plots represents the proportion of the microlensing alert presented to the model, ranging from 0\% (no microlensing signature shown) to 100\% (all the observations between $t_{break}$ and $t_{cut}$ are shown to the model, i.e., the full alerted region as identified by the AlertFinder algorithm). To obtain the modified datasets with the different percentages we performed various right-side cropping operations to the region of the time-series within $t_{break}$ and $t_{cut}$. This experiment provided valuable insights into how the model's performance evolves as it ``sees" less of the microlensing alert. 

It is important to note that in the operational mode, the network should function at 100\% alert visibility, as this is the data set provided by the AlertFinder algorithm. However, we conducted this test to explore the model's behavior under more constrained conditions, pushing its limits to better understand its robustness and reliability.

The left plots in the Fig. \ref{fig:fig_11_acc_per_percentshown} show the per-category and per-set accuracies. These plots represent the individual categorical accuracy for both the binary and multi-class classification tasks across the training, validation, and test sets.

The right plots of the Fig. \ref{fig:fig_11_acc_per_percentshown} present the mean accuracies for each set, calculated by averaging the curves of each category displayed in the left panel. These provide a consolidated view of the model's performance across the training, validation, and test sets for both the binary and multi-class classification tasks.

The categorical accuracy is calculated as the number of correctly classified instances in the category divided by the total number of instances in that category. Due to the class imbalance in the data, it was necessary to break down the accuracies of each category independently to ensure a more accurate assessment of the model's performance across all classes.

At 0\% microlensing alert visibility, the model operates essentially at chance level. Thus, it should approximately have an accuracy of 50\% for the binary classification task and 33\% for the three-class classification task. In the Fig. \ref{fig:fig_11_acc_per_percentshown} however we see that the aggregated accuracies on the right panel start at values of $\sim 55\% - 60\%$ and $\sim 40\% - 50\%$. In the cases we checked, this discrepancy can be attributed to instances where the microlensing event begins slightly earlier than the designated start time, yet the AlertFinder algorithm assigns a $t_{break}$ value that is marginally delayed.

In both tasks, a clear steadily-increasing upward trend is observed in the mean accuracy as the model sees more of the microlensing alert. In the binary classification task, the most significant gains occur between 0\% and 40\% of alert visibility, suggesting that even partial visibility of the microlensing event provides sufficient information for the model to make accurate predictions. However, in the multi-class task, the most notable improvements in accuracy occur in the final region, from 60\% to 100\% visibility. This effect, coupled with the concave shape of the mean accuracy curves of this second task, indicates that the model's accuracy significantly benefits from gaining more visibility of the alert in order to distinguish between the ``Yes" and ``Maybe" categories (as it can be seen in the bottom left plot of Fig. \ref{fig:fig_11_acc_per_percentshown} that the accuracy of the ``Yes" category increases significantly in this visibility region, while the accuracies of the ``No" and ``Maybe" categories do not have as much improvement). In contrast, the mean accuracy curves of the binary task exhibit a convex curve (just as the ``No" categorical accuracy curve does), which tell us that increased visibility is less critical. In other words, the model is already able to distinguish accurately between ``Yes"/``Maybe" and ``No" with only part of the microlensing data.

As expected, the mean accuracy of the models peak at 100\% alert visibility, with values of  $\sim 85\%$ and $\sim 78\%$ for the binary and multi-class tasks, respectively. This reflects the models' effectiveness in the real scenario where it has access to the complete alert data. Logically, the accuracy of the multi-class task is lower, highlighting the added complexity of differentiating between three categories. This result underscores the challenges of multi-class classification, where subtle distinctions between classes must be learned from the data. The overall trend, though, remains positive, showing that the model can effectively leverage the full alert data to enhance its predictions.

\subsection{LensNet Threshold Analysis} \label{sec:results_threshold}

Fig. \ref{fig:fig_12_accs_per_thres} illustrates the relationship between the threshold of the output neuron (on the horizontal axis) and the accuracy (on the vertical axis) for the binary classification task. In this context, the threshold determines the decision boundary for classifying instances into either category. 

The green curves represent the accuracy for the ``Yes" and ``Maybe" category across the different sets, while the red curves represent the accuracy for the ``No" category. The brown curves show the average accuracy across both categories (which in a binary classification problem should always begin and end at 50\%).

As the threshold increases, it requires a higher value from the output of the model in order to classify the instance as a ``Yes" or ``Maybe". In other words, the model needs higher confidence on the output so its classified in the ``Yes" or ``Maybe" category. 

The benefit of this behavior is that the ``Yes" or ``Maybe" predictions will be of high purity, at the expense of more "false negatives"; that is true microlensing events will be incorrectly classified as ``No" more frequently.

The peak of the mean accuracy curves (brown) indicates the most optimal operational threshold, which is $\sim 0.45$ (dashed black vertical line of the left). At this threshold, the model achieves a balanced performance between the two categories, resulting in an average accuracy of 87.5\% in the test set. This threshold coincides with the intersection of the green and red curves, representing the best trade-off between correctly classifying instances of both categories. This ensures that neither category's accuracy is disproportionately compromised, maximizing overall model performance.

In the curves shown in Figs. \ref{fig:fig_9_acc_evol_bin} and \ref{fig:fig_10_acc_evol_mult}, the final epoch accuracies are lower compared to those in Fig. \ref{fig:fig_12_accs_per_thres} due to the use of the augmented dataset during training, which posed a more challenging task for the model. In contrast, the accuracy curves in Fig. \ref{fig:fig_12_accs_per_thres} are calculated using the non-augmented data, resulting in higher performance values.

On the other hand, for our application, we prioritize maximizing the purity of actual microlensing events as opposed to having a balanced classification performance in both categories, which leads us to select a higher threshold. This is mainly due to the high relative costs of processing and storing false positives due to the high number of instances required to analyzed every day.

At a threshold of 0.955 (dashed black vertical line of the right), the model achieves $\sim 100\%$ accuracy in the ``No" category while retaining approximately 30\% of true ``Yes" instances. This high threshold is particularly important in minimizing false positives, aligning with our goal of maintaining high purity in the ``Yes"/``Maybe" predictions.

\section{Discussion} \label{sec:discuss}

In this work, we developed a machine learning model to classify microlensing events using data from the Korea Microlensing Telescope Network (KMTNet). Our model, based on a branched Recurrent Neural Network (RNN) architecture, was specifically designed to handle time-series data from multiple telescopes and features. The key focus of this study was to assess the model's ability to detect genuine microlensing events, including distinguishing between the ``Yes," ``Maybe," and ``No" categories. Through the use of data augmentation and careful preprocessing, the model was trained to handle varying levels of data visibility, which provided important insights into the robustness of the classification process.

\subsection{LensNet Performance} \label{sec:discuss_performance}

Our results showed that the model performs effectively across both the binary and 3-class classification tasks. In the binary task, the model achieved a rapid convergence to a stable performance, with a notable peak accuracy of around 85\% when trained and tested on non-augmented data. For the more complex 3-class task, the model displayed a longer learning curve, particularly when trying to distinguish between the ``Yes" and ``Maybe" categories, which are often difficult to differentiate. Nevertheless, the model's performance continued to improve as more data was made available, and it reached an accuracy of approximately 78\% when fully trained. The difference in the performance between the binary and multi-class tasks highlights the added complexity of distinguishing between more than two categories, particularly when ``Maybe" events are involved.

A key finding from our experiments was that even with partial visibility of the microlensing alert data, the model could make accurate predictions, especially in the binary task. However, for the 3-class classification, greater visibility was crucial to improving the accuracy for the ``Yes" category, especially in distinguishing between ``Yes" and ``Maybe" events. The results also revealed that the model can be fine-tuned using different thresholds, allowing us to prioritize either a balanced classification performance or a higher purity of microlensing event detection, depending on the application's requirements. This flexibility is particularly important in practical deployment scenarios, where false positives can be costly.

A particularly noteworthy achievement of our model is its ability to often distinguish ``Maybe" events from flux data alone, without relying on difference images. This is a significant advancement, especially considering that human experts traditionally depend on difference images to make such classifications. The difference image provides critical insights, such as the presence of diffraction spikes, pixel bleeding, or other instrumental anomalies that could artificially inflate flux values and mimic a microlensing event. By using these images, human vetters can effectively rule out false positives and identify true microlensing events with high confidence.

However, our model has demonstrated a good capability to differentiate between genuine microlensing events and false positives purely from time-series flux data. This ability is particularly important because it suggests that the model can identify subtle patterns and features within the flux data that are indicative of real microlensing events, even in the absence of additional diagnostic tools like difference images.

\subsection{Prospects for Deployment} \label{sec:deploy}

The integration of LensNet into real-time microlensing classification pipelines presents an opportunity for improvement of the current system of KMTNet, by significantly reducing the need for manual human vetting. Traditionally, human experts review thousands of alerts daily, a process fraught with inconsistencies and inefficiencies. By automating a substantial portion of this workload, our model should rapidly and accurately classify microlensing events directly from time-series data, allowing for faster decision-making and more efficient use of human resources.

Because we can select different thresholds we can choose to prioritize either high-purity, balanced or high recall classifications. For example, by setting a threshold of 0.955 we can achieve a highly pure sample with near-perfect accuracy for identifying non-events. This would be a useful choice for automating the AlertFinder because it would identify the highest-confidence events but not result in many false positives being alerted. Specifically for this high threshold configuration, LensNet achieves over 99.7\% accuracy in the ``No" category, ensuring that human reviewers are not overwhelmed by false positives. Although this higher threshold reduces the capture rate of true microlensing events to about 30-40\%, this trade-off is strategically advantageous. By minimizing false positives, the review process becomes significantly more efficient, allowing human experts to focus only on the most promising and relevant alerts. Nevertheless, the flexible threshold mechanism in the model allows for real-time adjustments depending on the operational needs. 

On the other hand, if we wanted to be more permissive with LensNet, allowing more positively classified events, we could choose a lower threshold (~0.45) which could be used for more balanced accuracy. This would result in higher false-positive rates. Therefore, we would need another filter before alerting the community to events identified in this preliminary stage. Indeed, LensNet replaces only one step in the vetting process. Following classification, candidates still undergo a secondary stage of difference imaging, where false positives can be further reduced through human review or automated methods \citep{deBeurs_inprep}. Nonetheless, even if we choose a more permissive approach with a lower threshold in LensNet, this secondary algorithm should allow us to remove the overhead of false positives. 

We are now investigating the deployment of the model into the KMTNet team's real-time classification systems. By integrating LensNet directly into the internal processing pipelines, it will deliver real-time classifications as alerts are generated. This automation will enable KMTNet to process large volumes of data more efficiently, allowing human reviewers to concentrate their efforts on the most promising events, significantly improving the overall workflow.

\subsection{Future Work} \label{sec:future_work}

Looking ahead, our future work will involve developing and training a secondary pipeline that integrates difference images into the model’s predictions. While our current approach focuses on time-series data due to its lower computational demands, incorporating difference images could further enhance the accuracy of our classifications, particularly in challenging cases. However, the process of generating and analyzing difference images is computationally expensive and requires substantial memory allocation, making it feasible only for cases with a high likelihood of being true microlensing events. By combining the strengths of flux data analysis with the diagnostic power of difference images, we aim to create a more robust and precise detection pipeline, capable of handling even the most complex scenarios in microlensing event classification.

\section{Conclusions} \label{sec:concl}

We have developed a model that represents a significant advancement in the automated classification of microlensing events, the deployment of which will fundamentally change the way these events are detected in the KMTNet. Historically, the reliance on human experts for manual vetting has constrained the scalability of microlensing surveys. By automating a significant portion of this process, the system can enable faster and more efficient follow-ups, allowing astronomers to quickly flag potential microlensing events and prioritize observations, thereby increasing exoplanet discoveries and the likelihood of capturing transient phenomena in real time. This advancement also facilitates the scaling of microlensing surveys beyond what is currently possible, allowing for the exploration of larger portions of the sky and deeper investigations into planetary systems around distant stars. 

The model, LensNet, utilizes a branched RNN architecture that integrates both time-series flux data and auxiliary features. It achieved high levels of accuracy, particularly in the binary classification task, demonstrating a strong ability to generalize across varying levels of alert visibility. While the 3-class classification task remains more challenging, LensNet has shown consistent improvements as more data become available, offering promise for enhancing multi-class detection performance in real-time applications. Additionally, its flexibility in adjusting output neuron threshold allows LensNet to achieve higher purity in the classification of real microlensing events, making it especially useful for minimizing false positives in large-scale surveys. 
Future work will focus on incorporating additional data sources, such as difference images, to further refine the model’s accuracy and reliability in complex scenarios. 

 While developed for KMTNet, our approach can be adapted and applied to upcoming missions such as the Nancy Grace Roman Space Telescope. In particular, the Roman Galactic Bulge Time Domain Survey \citep{Gaudi2022} is expected to detect more than $30,000$ microlensing events \citep{Penny2019}. Ultimately, LensNet has the potential to revolutionize microlensing event detection, paving the way for more efficient, large-scale exploration of the universe.

\begin{acknowledgements}
This research has made use of the KMTNet system
operated by the Korea Astronomy and Space Science Institute
(KASI) at three host sites of CTIO in Chile, SAAO in South
Africa, and SSO in Australia. Data transfer from the host site to
KASI was supported by the Korea Research Environment
Open NETwork (KREONET). This research was supported by KASI
under the R\&D program (project No. 2024-1-832-01) supervised
by the Ministry of Science and ICT.

J.V. acknowledges support from the Mauricio and Carlota Botton Foundation.

J.C.Y. and I.-G.S. acknowledge support from U.S. NSF Grant No. AST-2108414. 

W.Zang, H.Y., S.M., R.K., J.Z., and W.Zhu acknowledge support by the National Natural Science Foundation of China (Grant No. 12133005). 

W.Zang acknowledges the support from the Harvard-Smithsonian Center for Astrophysics through the CfA Fellowship. 

Y.S. acknowledges support from BSF Grant No. 2020740.

Work by C.H. was supported by the grants of National Research Foundation of Korea (2019R1A2C2085965 and 2020R1A4A2002885). 
 
\end{acknowledgements}

\bibliography{references}{}

\begin{thebibliography}{}
\expandafter\ifx\csname natexlab\endcsname\relax\def\natexlab#1{#1}\fi
\providecommand{\url}[1]{\href{#1}{#1}}
\providecommand{\dodoi}[1]{doi:~\href{http://doi.org/#1}{\nolinkurl{#1}}}
\providecommand{\doeprint}[1]{\href{http://ascl.net/#1}{\nolinkurl{http://ascl.net/#1}}}
\providecommand{\doarXiv}[1]{\href{https://arxiv.org/abs/#1}{\nolinkurl{https://arxiv.org/abs/#1}}}

\bibitem[{Abadi {et~al.}(2015)Abadi, Agarwal, Barham, Brevdo, Chen, Citro, Corrado, Davis, Dean, Devin, Ghemawat, Goodfellow, Harp, Irving, Isard, Jia, Jozefowicz, Kaiser, Kudlur, Levenberg, Man\'{e}, Monga, Moore, Murray, Olah, Schuster, Shlens, Steiner, Sutskever, Talwar, Tucker, Vanhoucke, Vasudevan, Vi\'{e}gas, Vinyals, Warden, Wattenberg, Wicke, Yu, \& Zheng}]{tensorflow2015-whitepaper}
Abadi, M., Agarwal, A., Barham, P., {et~al.} 2015, {TensorFlow}: Large-Scale Machine Learning on Heterogeneous Systems.
\newblock \url{https://www.tensorflow.org/}

\bibitem[{{Abe} {et~al.}(2013){Abe}, {Airey}, {Barnard}, {Baudry}, {Botzler}, {Douchin}, {Freeman}, {Larsen}, {Niemiec}, {Perrott}, {Philpott}, {Rattenbury}, \& {Yock}}]{Abe13}
{Abe}, F., {Airey}, C., {Barnard}, E., {et~al.} 2013, \mnras, 431, 2975, \dodoi{10.1093/mnras/stt318}

\bibitem[{{Abrams} {et~al.}(2023){Abrams}, {Hundertmark}, {Khakpash}, {Street}, {Jones}, {Lu}, {Bachelet}, {Tsapras}, {Moniez}, {Blaineauu}, {Di Stefano}, {Makler}, {Varela}, \& {Rabus}}]{Abrams23_LSST}
{Abrams}, N.~S., {Hundertmark}, M. P.~G., {Khakpash}, S., {et~al.} 2023, arXiv e-prints, arXiv:2309.15310, \dodoi{10.48550/arXiv.2309.15310}

\bibitem[{{Alcock} {et~al.}(1996){Alcock}, {Allsman}, {Axelrod}, {Bennett}, {Cook}, {Freeman}, {Griest}, {Guern}, {Lehner}, {Marshall}, {Park}, {Perlmutter}, {Peterson}, {Pratt}, {Quinn}, {Rodgers}, {Stubbs}, \& {Sutherland}}]{Alcock1996}
{Alcock}, C., {Allsman}, R.~A., {Axelrod}, T.~S., {et~al.} 1996, \apj, 461, 84, \dodoi{10.1086/177039}

\bibitem[{{Beaulieu} {et~al.}(2006){Beaulieu}, {Bennett}, {Fouqu{\'e}}, {Williams}, {Dominik}, {J{\o}rgensen}, {Kubas}, {Cassan}, {Coutures}, {Greenhill}, {Hill}, {Menzies}, {Sackett}, {Albrow}, {Brillant}, {Caldwell}, {Calitz}, {Cook}, {Corrales}, {Desort}, {Dieters}, {Dominis}, {Donatowicz}, {Hoffman}, {Kane}, {Marquette}, {Martin}, {Meintjes}, {Pollard}, {Sahu}, {Vinter}, {Wambsganss}, {Woller}, {Horne}, {Steele}, {Bramich}, {Burgdorf}, {Snodgrass}, {Bode}, {Udalski}, {Szyma{\'n}ski}, {Kubiak}, {Wi{\c e}ckowski}, {Pietrzy{\'n}ski}, {Soszy{\'n}ski}, {Szewczyk}, {Wyrzykowski}, {Paczy{\'n}ski}, {Abe}, {Bond}, {Britton}, {Gilmore}, {Hearnshaw}, {Itow}, {Kamiya}, {Kilmartin}, {Korpela}, {Masuda}, {Matsubara}, {Motomura}, {Muraki}, {Nakamura}, {Okada}, {Ohnishi}, {Rattenbury}, {Sako}, {Sato}, {Sasaki}, {Sekiguchi}, {Sullivan}, {Tristram}, {Yock}, \& {Yoshioka}}]{Beaulieu06}
{Beaulieu}, J.-P., {Bennett}, D.~P., {Fouqu{\'e}}, P., {et~al.} 2006, \nat, 439, 437, \dodoi{10.1038/nature04441}

\bibitem[{{Bellm} {et~al.}(2019){Bellm}, {Kulkarni}, {Graham}, {Dekany}, {Smith}, {Riddle}, {Masci}, {Helou}, {Prince}, {Adams}, {Barbarino}, {Barlow}, {Bauer}, {Beck}, {Belicki}, {Biswas}, {Blagorodnova}, {Bodewits}, {Bolin}, {Brinnel}, {Brooke}, {Bue}, {Bulla}, {Burruss}, {Cenko}, {Chang}, {Connolly}, {Coughlin}, {Cromer}, {Cunningham}, {De}, {Delacroix}, {Desai}, {Duev}, {Eadie}, {Farnham}, {Feeney}, {Feindt}, {Flynn}, {Franckowiak}, {Frederick}, {Fremling}, {Gal-Yam}, {Gezari}, {Giomi}, {Goldstein}, {Golkhou}, {Goobar}, {Groom}, {Hacopians}, {Hale}, {Henning}, {Ho}, {Hover}, {Howell}, {Hung}, {Huppenkothen}, {Imel}, {Ip}, {Ivezi{\'c}}, {Jackson}, {Jones}, {Juric}, {Kasliwal}, {Kaspi}, {Kaye}, {Kelley}, {Kowalski}, {Kramer}, {Kupfer}, {Landry}, {Laher}, {Lee}, {Lin}, {Lin}, {Lunnan}, {Giomi}, {Mahabal}, {Mao}, {Miller}, {Monkewitz}, {Murphy}, {Ngeow}, {Nordin}, {Nugent}, {Ofek}, {Patterson}, {Penprase}, {Porter}, {Rauch}, {Rebbapragada}, {Reiley}, {Rigault}, {Rodriguez}, {van Roestel}, {Rusholme}, {van
  Santen}, {Schulze}, {Shupe}, {Singer}, {Soumagnac}, {Stein}, {Surace}, {Sollerman}, {Szkody}, {Taddia}, {Terek}, {Van Sistine}, {van Velzen}, {Vestrand}, {Walters}, {Ward}, {Ye}, {Yu}, {Yan}, \& {Zolkower}}]{Bellm19_ZTF}
{Bellm}, E.~C., {Kulkarni}, S.~R., {Graham}, M.~J., {et~al.} 2019, \pasp, 131, 018002, \dodoi{10.1088/1538-3873/aaecbe}

\bibitem[{{Bensby} {et~al.}(2013){Bensby}, {Yee}, {Feltzing}, {Johnson}, {Gould}, {Cohen}, {Asplund}, {Mel{\'e}ndez}, {Lucatello}, {Han}, {Thompson}, {Gal-Yam}, {Udalski}, {Bennett}, {Bond}, {Kohei}, {Sumi}, {Suzuki}, {Suzuki}, {Takino}, {Tristram}, {Yamai}, \& {Yonehara}}]{Bensby13}
{Bensby}, T., {Yee}, J.~C., {Feltzing}, S., {et~al.} 2013, \aap, 549, A147, \dodoi{10.1051/0004-6361/201220678}

\bibitem[{{Bond} {et~al.}(2001){Bond}, {Abe}, {Dodd}, {Hearnshaw}, {Honda}, {Jugaku}, {Kilmartin}, {Marles}, {Masuda}, {Matsubara}, {Muraki}, {Nakamura}, {Nankivell}, {Noda}, {Noguchi}, {Ohnishi}, {Rattenbury}, {Reid}, {Saito}, {Sato}, {Sekiguchi}, {Skuljan}, {Sullivan}, {Sumi}, {Takeuti}, {Watase}, {Wilkinson}, {Yamada}, {Yanagisawa}, \& {Yock}}]{Bond01}
{Bond}, I.~A., {Abe}, F., {Dodd}, R.~J., {et~al.} 2001, \mnras, 327, 868, \dodoi{10.1046/j.1365-8711.2001.04776.x}

\bibitem[{{Bond} {et~al.}(2004){Bond}, {Udalski}, {Jaroszy{\'n}ski}, {Rattenbury}, {Paczy{\'n}ski}, {Soszy{\'n}ski}, {Wyrzykowski}, {Szyma{\'n}ski}, {Kubiak}, {Szewczyk}, {{\.Z}ebru{\'n}}, {Pietrzy{\'n}ski}, {Abe}, {Bennett}, {Eguchi}, {Furuta}, {Hearnshaw}, {Kamiya}, {Kilmartin}, {Kurata}, {Masuda}, {Matsubara}, {Muraki}, {Noda}, {Okajima}, {Sako}, {Sekiguchi}, {Sullivan}, {Sumi}, {Tristram}, {Yanagisawa}, \& {Yock}}]{Bond04}
{Bond}, I.~A., {Udalski}, A., {Jaroszy{\'n}ski}, M., {et~al.} 2004, \apjl, 606, L155, \dodoi{10.1086/420928}

\bibitem[{{Boone}(2019)}]{Boone19_Avocado}
{Boone}, K. 2019, \aj, 158, 257, \dodoi{10.3847/1538-3881/ab5182}

\bibitem[{{Chu} {et~al.}(2019){Chu}, {Wagstaff}, {Bryden}, \& {Shvartzvald}}]{Chu19}
{Chu}, S., {Wagstaff}, K., {Bryden}, G., \& {Shvartzvald}, Y. 2019, in Astronomical Society of the Pacific Conference Series, Vol. 523, Astronomical Data Analysis Software and Systems XXVII, ed. P.~J. {Teuben}, M.~W. {Pound}, B.~A. {Thomas}, \& E.~M. {Warner}, 127

\bibitem[{{de Beurs}(in prep)}]{deBeurs_inprep}
{de Beurs}, Zo\:{e}, e. in prep, in prep

\bibitem[{{Dong} {et~al.}(2019){Dong}, {M{\'e}rand}, {Delplancke-Str{\"o}bele}, {Gould}, {Chen}, {Post}, {Kochanek}, {Stanek}, {Christie}, {Mutel}, {Natusch}, {Holoien}, {Prieto}, {Shappee}, \& {Thompson}}]{Dong19_GRAVITY}
{Dong}, S., {M{\'e}rand}, A., {Delplancke-Str{\"o}bele}, F., {et~al.} 2019, \apj, 871, 70, \dodoi{10.3847/1538-4357/aaeffb}

\bibitem[{{Gaudi}(2022)}]{Gaudi2022}
{Gaudi}, B.~S. 2022, in Bulletin of the American Astronomical Society, Vol.~54, 102.146

\bibitem[{{Gezer} {et~al.}(2022){Gezer}, {Wyrzykowski}, {Zieli{\'n}ski}, {Marton}, {Kruszy{\'n}ska}, {Rybicki}, {Ihanec}, {Jab{\l}o{\'n}ska}, \& {Zi{\'o}{\l}kowska}}]{Gezer22}
{Gezer}, I., {Wyrzykowski}, {\L}., {Zieli{\'n}ski}, P., {et~al.} 2022, arXiv e-prints, arXiv:2201.12209, \dodoi{10.48550/arXiv.2201.12209}

\bibitem[{{Godines} {et~al.}(2019){Godines}, {Bachelet}, {Narayan}, \& {Street}}]{Godines19}
{Godines}, D., {Bachelet}, E., {Narayan}, G., \& {Street}, R.~A. 2019, Astronomy and Computing, 28, 100298, \dodoi{10.1016/j.ascom.2019.100298}

\bibitem[{{Gould}(2013)}]{Gould13LSST}
{Gould}, A. 2013, ArXiv e-prints.
\newblock \doarXiv{1304.3455}

\bibitem[{{Gould} \& {Loeb}(1992)}]{GouldLoeb92}
{Gould}, A., \& {Loeb}, A. 1992, \apj, 396, 104, \dodoi{10.1086/171700}

\bibitem[{{Graham} {et~al.}(2019){Graham}, {Kulkarni}, {Bellm}, {Adams}, {Barbarino}, {Blagorodnova}, {Bodewits}, {Bolin}, {Brady}, {Cenko}, {Chang}, {Coughlin}, {De}, {Eadie}, {Farnham}, {Feindt}, {Franckowiak}, {Fremling}, {Gezari}, {Ghosh}, {Goldstein}, {Golkhou}, {Goobar}, {Ho}, {Huppenkothen}, {Ivezi{\'c}}, {Jones}, {Juric}, {Kaplan}, {Kasliwal}, {Kelley}, {Kupfer}, {Lee}, {Lin}, {Lunnan}, {Mahabal}, {Miller}, {Ngeow}, {Nugent}, {Ofek}, {Prince}, {Rauch}, {van Roestel}, {Schulze}, {Singer}, {Sollerman}, {Taddia}, {Yan}, {Ye}, {Yu}, {Barlow}, {Bauer}, {Beck}, {Belicki}, {Biswas}, {Brinnel}, {Brooke}, {Bue}, {Bulla}, {Burruss}, {Connolly}, {Cromer}, {Cunningham}, {Dekany}, {Delacroix}, {Desai}, {Duev}, {Feeney}, {Flynn}, {Frederick}, {Gal-Yam}, {Giomi}, {Groom}, {Hacopians}, {Hale}, {Helou}, {Henning}, {Hover}, {Hillenbrand}, {Howell}, {Hung}, {Imel}, {Ip}, {Jackson}, {Kaspi}, {Kaye}, {Kowalski}, {Kramer}, {Kuhn}, {Landry}, {Laher}, {Mao}, {Masci}, {Monkewitz}, {Murphy}, {Nordin}, {Patterson}, {Penprase},
  {Porter}, {Rebbapragada}, {Reiley}, {Riddle}, {Rigault}, {Rodriguez}, {Rusholme}, {van Santen}, {Shupe}, {Smith}, {Soumagnac}, {Stein}, {Surace}, {Szkody}, {Terek}, {Van Sistine}, {van Velzen}, {Vestrand}, {Walters}, {Ward}, {Zhang}, \& {Zolkower}}]{Graham19_ZTF}
{Graham}, M.~J., {Kulkarni}, S.~R., {Bellm}, E.~C., {et~al.} 2019, \pasp, 131, 078001, \dodoi{10.1088/1538-3873/ab006c}

\bibitem[{{Kim} {et~al.}(2018{\natexlab{a}}){Kim}, {Kim}, {Hwang}, {Albrow}, {Chung}, {Gould}, {Han}, {Jung}, {Ryu}, {Shin}, {Yee}, {Zhu}, {Cha}, {Kim}, {Lee}, {Lee}, {Lee}, {Park}, {Pogge}, \& {KMTNet Collaboration}}]{KimKim18_EF}
{Kim}, D.~J., {Kim}, H.~W., {Hwang}, K.~H., {et~al.} 2018{\natexlab{a}}, \aj, 155, 76, \dodoi{10.3847/1538-3881/aaa47b}

\bibitem[{{Kim} {et~al.}(2018{\natexlab{b}}){Kim}, {Hwang}, {Shvartzvald}, {Yee}, {Albrow}, {Cha}, {Chung}, {Gould}, {Han}, {Jung}, {Kim}, {Kim}, {Lee}, {Lee}, {Lee}, {Park}, {Pogge}, {Ryu}, {Shin}, \& {Zang}}]{Kim18_AF}
{Kim}, H.-W., {Hwang}, K.-H., {Shvartzvald}, Y., {et~al.} 2018{\natexlab{b}}, arXiv e-prints, arXiv:1806.07545.
\newblock \doarXiv{1806.07545}

\bibitem[{{Kim} {et~al.}(2018{\natexlab{c}}){Kim}, {Hwang}, {Kim}, {Albrow}, {Cha}, {Chung}, {Gould}, {Han}, {Jung}, {Kim}, {Lee}, {Lee}, {Lee}, {Park}, {Pogge}, {Ryu}, {Shin}, {\~{}Shvartzvald}, {Yee}, {Zang}, \& {Zhu}}]{Kim18EF}
{Kim}, H.-W., {Hwang}, K.-H., {Kim}, D.-J., {et~al.} 2018{\natexlab{c}}, ArXiv e-prints.
\newblock \doarXiv{1804.03352}

\bibitem[{{Kim} {et~al.}(2016){Kim}, {Lee}, {Park}, {Kim}, {Cha}, {Lee}, {Han}, {Chun}, \& {Yuk}}]{Kim2016}
{Kim}, S.-L., {Lee}, C.-U., {Park}, B.-G., {et~al.} 2016, Journal of Korean Astronomical Society, 49, 37, \dodoi{10.5303/JKAS.2016.49.1.37}

\bibitem[{Kingma(2014)}]{kingma2014adam}
Kingma, D.~P. 2014, arXiv preprint arXiv:1412.6980

\bibitem[{{Kochanek} {et~al.}(2017){Kochanek}, {Shappee}, {Stanek}, {Holoien}, {Thompson}, {Prieto}, {Dong}, {Shields}, {Will}, {Britt}, {Perzanowski}, \& {Pojma{\'n}ski}}]{Kochanek17_ASAS}
{Kochanek}, C.~S., {Shappee}, B.~J., {Stanek}, K.~Z., {et~al.} 2017, \pasp, 129, 104502, \dodoi{10.1088/1538-3873/aa80d9}

\bibitem[{{Lam} {et~al.}(2022){Lam}, {Lu}, {Udalski}, {Bond}, {Bennett}, {Skowron}, {Mr{\'o}z}, {Poleski}, {Sumi}, {Szyma{\'n}ski}, {Koz{\l}owski}, {Pietrukowicz}, {Soszy{\'n}ski}, {Ulaczyk}, {Wyrzykowski}, {Miyazaki}, {Suzuki}, {Koshimoto}, {Rattenbury}, {Hosek}, {Abe}, {Barry}, {Bhattacharya}, {Fukui}, {Fujii}, {Hirao}, {Itow}, {Kirikawa}, {Kondo}, {Matsubara}, {Matsumoto}, {Muraki}, {Olmschenk}, {Ranc}, {Okamura}, {Satoh}, {Silva}, {Toda}, {Tristram}, {Vandorou}, {Yama}, {Abrams}, {Agarwal}, {Rose}, \& {Terry}}]{Lam22_BH}
{Lam}, C.~Y., {Lu}, J.~R., {Udalski}, A., {et~al.} 2022, \apjl, 933, L23, \dodoi{10.3847/2041-8213/ac7442}

\bibitem[{{LSST Science Collaboration} {et~al.}(2009){LSST Science Collaboration}, {Abell}, {Allison}, {Anderson}, {Andrew}, {Angel}, {Armus}, {Arnett}, {Asztalos}, {Axelrod}, {Bailey}, {Ballantyne}, {Bankert}, {Barkhouse}, {Barr}, {Barrientos}, {Barth}, {Bartlett}, {Becker}, {Becla}, {Beers}, {Bernstein}, {Biswas}, {Blanton}, {Bloom}, {Bochanski}, {Boeshaar}, {Borne}, {Bradac}, {Brandt}, {Bridge}, {Brown}, {Brunner}, {Bullock}, {Burgasser}, {Burge}, {Burke}, {Cargile}, {Chandrasekharan}, {Chartas}, {Chesley}, {Chu}, {Cinabro}, {Claire}, {Claver}, {Clowe}, {Connolly}, {Cook}, {Cooke}, {Cooray}, {Covey}, {Culliton}, {de Jong}, {de Vries}, {Debattista}, {Delgado}, {Dell'Antonio}, {Dhital}, {Di Stefano}, {Dickinson}, {Dilday}, {Djorgovski}, {Dobler}, {Donalek}, {Dubois-Felsmann}, {Durech}, {Eliasdottir}, {Eracleous}, {Eyer}, {Falco}, {Fan}, {Fassnacht}, {Ferguson}, {Fernandez}, {Fields}, {Finkbeiner}, {Figueroa}, {Fox}, {Francke}, {Frank}, {Frieman}, {Fromenteau}, {Furqan}, {Galaz}, {Gal-Yam}, {Garnavich},
  {Gawiser}, {Geary}, {Gee}, {Gibson}, {Gilmore}, {Grace}, {Green}, {Gressler}, {Grillmair}, {Habib}, {Haggerty}, {Hamuy}, {Harris}, {Hawley}, {Heavens}, {Hebb}, {Henry}, {Hileman}, {Hilton}, {Hoadley}, {Holberg}, {Holman}, {Howell}, {Infante}, {Ivezic}, {Jacoby}, {Jain}, {R}, {Jedicke}, {Jee}, {Garrett Jernigan}, {Jha}, {Johnston}, {Jones}, {Juric}, {Kaasalainen}, {Styliani}, {Kafka}, {Kahn}, {Kaib}, {Kalirai}, {Kantor}, {Kasliwal}, {Keeton}, {Kessler}, {Knezevic}, {Kowalski}, {Krabbendam}, {Krughoff}, {Kulkarni}, {Kuhlman}, {Lacy}, {Lepine}, {Liang}, {Lien}, {Lira}, {Long}, {Lorenz}, {Lotz}, {Lupton}, {Lutz}, {Macri}, {Mahabal}, {Mandelbaum}, {Marshall}, {May}, {McGehee}, {Meadows}, {Meert}, {Milani}, {Miller}, {Miller}, {Mills}, {Minniti}, {Monet}, {Mukadam}, {Nakar}, {Neill}, {Newman}, {Nikolaev}, {Nordby}, {O'Connor}, {Oguri}, {Oliver}, {Olivier}, {Olsen}, {Olsen}, {Olszewski}, {Oluseyi}, {Padilla}, {Parker}, {Pepper}, {Peterson}, {Petry}, {Pinto}, {Pizagno}, {Popescu}, {Prsa}, {Radcka}, {Raddick},
  {Rasmussen}, {Rau}, {Rho}, {Rhoads}, {Richards}, {Ridgway}, {Robertson}, {Roskar}, {Saha}, {Sarajedini}, {Scannapieco}, {Schalk}, {Schindler}, {Schmidt}, {Schmidt}, {Schneider}, {Schumacher}, {Scranton}, {Sebag}, {Seppala}, {Shemmer}, {Simon}, {Sivertz}, {Smith}, {Allyn Smith}, {Smith}, {Spitz}, {Stanford}, {Stassun}, {Strader}, {Strauss}, {Stubbs}, {Sweeney}, {Szalay}, {Szkody}, {Takada}, {Thorman}, {Trilling}, {Trimble}, {Tyson}, {Van Berg}, {Vanden Berk}, {VanderPlas}, {Verde}, {Vrsnak}, {Walkowicz}, {Wandelt}, {Wang}, {Wang}, {Warner}, {Wechsler}, {West}, {Wiecha}, {Williams}, {Willman}, {Wittman}, {Wolff}, {Wood-Vasey}, {Wozniak}, {Young}, {Zentner}, \& {Zhan}}]{LSST09}
{LSST Science Collaboration}, {Abell}, P.~A., {Allison}, J., {et~al.} 2009, arXiv e-prints, arXiv:0912.0201, \dodoi{10.48550/arXiv.0912.0201}

\bibitem[{{LSST Science Collaboration} {et~al.}(2017){LSST Science Collaboration}, {Marshall}, {Anguita}, {Bianco}, {Bellm}, {Brandt}, {Clarkson}, {Connolly}, {Gawiser}, {Ivezic}, {Jones}, {Lochner}, {Lund}, {Mahabal}, {Nidever}, {Olsen}, {Ridgway}, {Rhodes}, {Shemmer}, {Trilling}, {Vivas}, {Walkowicz}, {Willman}, {Yoachim}, {Anderson}, {Antilogus}, {Angus}, {Arcavi}, {Awan}, {Biswas}, {Bell}, {Bennett}, {Britt}, {Buzasi}, {Casetti-Dinescu}, {Chomiuk}, {Claver}, {Cook}, {Davenport}, {Debattista}, {Digel}, {Doctor}, {Firth}, {Foley}, {Fong}, {Galbany}, {Giampapa}, {Gizis}, {Graham}, {Grillmair}, {Gris}, {Haiman}, {Hartigan}, {Hawley}, {Hlozek}, {Jha}, {Johns-Krull}, {Kanbur}, {Kalogera}, {Kashyap}, {Kasliwal}, {Kessler}, {Kim}, {Kurczynski}, {Lahav}, {Liu}, {Malz}, {Margutti}, {Matheson}, {McEwen}, {McGehee}, {Meibom}, {Meyers}, {Monet}, {Neilsen}, {Newman}, {O'Dowd}, {Peiris}, {Penny}, {Peters}, {Poleski}, {Ponder}, {Richards}, {Rho}, {Rubin}, {Schmidt}, {Schuhmann}, {Shporer}, {Slater}, {Smith},
  {Soares-Santos}, {Stassun}, {Strader}, {Strauss}, {Street}, {Stubbs}, {Sullivan}, {Szkody}, {Trimble}, {Tyson}, {de Val-Borro}, {Valenti}, {Wagoner}, {Wood-Vasey}, \& {Zauderer}}]{LSST17}
{LSST Science Collaboration}, {Marshall}, P., {Anguita}, T., {et~al.} 2017, arXiv e-prints, arXiv:1708.04058, \dodoi{10.48550/arXiv.1708.04058}

\bibitem[{Mao {et~al.}(2023)Mao, Mohri, \& Zhong}]{mao2023cross}
Mao, A., Mohri, M., \& Zhong, Y. 2023, in International conference on Machine learning, PMLR, 23803--23828

\bibitem[{{Masci} {et~al.}(2019){Masci}, {Laher}, {Rusholme}, {Shupe}, {Groom}, {Surace}, {Jackson}, {Monkewitz}, {Beck}, {Flynn}, {Terek}, {Landry}, {Hacopians}, {Desai}, {Howell}, {Brooke}, {Imel}, {Wachter}, {Ye}, {Lin}, {Cenko}, {Cunningham}, {Rebbapragada}, {Bue}, {Miller}, {Mahabal}, {Bellm}, {Patterson}, {Juri{\'c}}, {Golkhou}, {Ofek}, {Walters}, {Graham}, {Kasliwal}, {Dekany}, {Kupfer}, {Burdge}, {Cannella}, {Barlow}, {Van Sistine}, {Giomi}, {Fremling}, {Blagorodnova}, {Levitan}, {Riddle}, {Smith}, {Helou}, {Prince}, \& {Kulkarni}}]{Masci19_ZTF}
{Masci}, F.~J., {Laher}, R.~R., {Rusholme}, B., {et~al.} 2019, \pasp, 131, 018003, \dodoi{10.1088/1538-3873/aae8ac}

\bibitem[{{Mr{\'o}z}(2020)}]{Mroz20_NN}
{Mr{\'o}z}, P. 2020, \actaa, 70, 169, \dodoi{10.32023/0001-5237/70.3.1}

\bibitem[{{Mr{\'o}z} {et~al.}(2022){Mr{\'o}z}, {Udalski}, \& {Gould}}]{Mroz22}
{Mr{\'o}z}, P., {Udalski}, A., \& {Gould}, A. 2022, \apjl, 937, L24, \dodoi{10.3847/2041-8213/ac90bb}

\bibitem[{{Nucita} {et~al.}(2018){Nucita}, {Licchelli}, {De Paolis}, {Ingrosso}, {Strafella}, {Katysheva}, \& {Shugarov}}]{Nucita18}
{Nucita}, A.~A., {Licchelli}, D., {De Paolis}, F., {et~al.} 2018, ArXiv e-prints.
\newblock \doarXiv{1802.06659}

\bibitem[{{Paczynski}(1986)}]{Paczynski86b}
{Paczynski}, B. 1986, \apj, 304, 1, \dodoi{10.1086/164140}

\bibitem[{{Penny} {et~al.}(2019){Penny}, {Gaudi}, {Kerins}, {Rattenbury}, {Mao}, {Robin}, \& {Calchi Novati}}]{Penny2019}
{Penny}, M.~T., {Gaudi}, B.~S., {Kerins}, E., {et~al.} 2019, \apjs, 241, 3, \dodoi{10.3847/1538-4365/aafb69}

\bibitem[{{Rybicki} {et~al.}(2022){Rybicki}, {Wyrzykowski}, {Bachelet}, {Cassan}, {Zieli{\'n}ski}, {Gould}, {Calchi Novati}, {Yee}, {Ryu}, {Gromadzki}, {Miko{\l}ajczyk}, {Ihanec}, {Kruszy{\'n}ska}, {Hambsch}, {Zo{\l}a}, {Fossey}, {Awiphan}, {Nakharutai}, {Lewis}, {Olivares E.}, {Hodgkin}, {Delgado}, {Breedt}, {Harrison}, {van Leeuwen}, {Rixon}, {Wevers}, {Yoldas}, {Udalski}, {Szyma{\'n}ski}, {Soszy{\'n}ski}, {Pietrukowicz}, {Koz{\l}owski}, {Skowron}, {Poleski}, {Ulaczyk}, {Mr{\'o}z}, {Iwanek}, {Wrona}, {Street}, {Tsapras}, {Hundertmark}, {Dominik}, {Beichman}, {Bryden}, {Carey}, {Gaudi}, {Henderson}, {Shvartzvald}, {Zang}, {Zhu}, {Christie}, {Green}, {Hennerley}, {McCormick}, {Monard}, {Natusch}, {Pogge}, {Gezer}, {Gurgul}, {Kaczmarek}, {Konacki}, {Lam}, {Maskoliunas}, {Pakstiene}, {Ratajczak}, {Stankeviciute}, {Zdanavicius}, \& {Zi{\'o}{\l}kowska}}]{Rybicki22_Gaia19bld}
{Rybicki}, K.~A., {Wyrzykowski}, {\L}., {Bachelet}, E., {et~al.} 2022, \aap, 657, A18, \dodoi{10.1051/0004-6361/202039542}

\bibitem[{{Sahu} {et~al.}(2022){Sahu}, {Anderson}, {Casertano}, {Bond}, {Udalski}, {Dominik}, {Calamida}, {Bellini}, {Brown}, {Rejkuba}, {Bajaj}, {Kains}, {Ferguson}, {Fryer}, {Yock}, {Mr{\'o}z}, {Koz{\l}owski}, {Pietrukowicz}, {Poleski}, {Skowron}, {Soszy{\'n}ski}, {Szyma{\'n}ski}, {Ulaczyk}, {Wyrzykowski}, {Barry}, {Bennett}, {Bond}, {Hirao}, {Silva}, {Kondo}, {Koshimoto}, {Ranc}, {Rattenbury}, {Sumi}, {Suzuki}, {Tristram}, {Vandorou}, {Beaulieu}, {Marquette}, {Cole}, {Fouqu{\'e}}, {Hill}, {Dieters}, {Coutures}, {Dominis-Prester}, {Bennett}, {Bachelet}, {Menzies}, {Albrow}, {Pollard}, {Gould}, {Yee}, {Allen}, {Almeida}, {Christie}, {Drummond}, {Gal-Yam}, {Gorbikov}, {Jablonski}, {Lee}, {Maoz}, {Manulis}, {McCormick}, {Natusch}, {Pogge}, {Shvartzvald}, {J{\o}rgensen}, {Alsubai}, {Andersen}, {Bozza}, {Novati}, {Burgdorf}, {Hinse}, {Hundertmark}, {Husser}, {Kerins}, {Longa-Pe{\~n}a}, {Mancini}, {Penny}, {Rahvar}, {Ricci}, {Sajadian}, {Skottfelt}, {Snodgrass}, {Southworth}, {Tregloan-Reed}, {Wambsganss},
  {Wertz}, {Tsapras}, {Street}, {Bramich}, {Horne}, {Steele}, \& {RoboNet Collaboration}}]{Sahu22_BH}
{Sahu}, K.~C., {Anderson}, J., {Casertano}, S., {et~al.} 2022, \apj, 933, 83, \dodoi{10.3847/1538-4357/ac739e}

\bibitem[{{Shappee} {et~al.}(2014){Shappee}, {Prieto}, {Grupe}, {Kochanek}, {Stanek}, {De Rosa}, {Mathur}, {Zu}, {Peterson}, {Pogge}, {Komossa}, {Im}, {Jencson}, {Holoien}, {Basu}, {Beacom}, {Szczygie{\l}}, {Brimacombe}, {Adams}, {Campillay}, {Choi}, {Contreras}, {Dietrich}, {Dubberley}, {Elphick}, {Foale}, {Giustini}, {Gonzalez}, {Hawkins}, {Howell}, {Hsiao}, {Koss}, {Leighly}, {Morrell}, {Mudd}, {Mullins}, {Nugent}, {Parrent}, {Phillips}, {Pojmanski}, {Rosing}, {Ross}, {Sand}, {Terndrup}, {Valenti}, {Walker}, \& {Yoon}}]{Shappee14_ASAS}
{Shappee}, B.~J., {Prieto}, J.~L., {Grupe}, D., {et~al.} 2014, \apj, 788, 48, \dodoi{10.1088/0004-637X/788/1/48}

\bibitem[{{Shvartzvald} {et~al.}(2016){Shvartzvald}, {Maoz}, {Udalski}, {Sumi}, {Friedmann}, {Kaspi}, {Poleski}, {Szyma{\'n}ski}, {Skowron}, {Koz{\l}owski}, {Wyrzykowski}, {Mr{\'o}z}, {Pietrukowicz}, {Pietrzy{\'n}ski}, {Soszy{\'n}ski}, {Ulaczyk}, {Abe}, {Barry}, {Bennett}, {Bhattacharya}, {Bond}, {Freeman}, {Inayama}, {Itow}, {Koshimoto}, {Ling}, {Masuda}, {Fukui}, {Matsubara}, {Muraki}, {Ohnishi}, {Rattenbury}, {Saito}, {Sullivan}, {Suzuki}, {Tristram}, {Wakiyama}, \& {Yonehara}}]{Shvartzvald16}
{Shvartzvald}, Y., {Maoz}, D., {Udalski}, A., {et~al.} 2016, \mnras, 457, 4089, \dodoi{10.1093/mnras/stw191}

\bibitem[{{Street} {et~al.}(2023){Street}, {Gough-Kelly}, {Lam}, {Varela}, {Makler}, {Bachelet}, {Lu}, {Abrams}, {Pusack}, {Terry}, {Di Stefano}, {Tsapras}, {Hundertmark}, {Grand}, {Daylan}, \& {Sobeck}}]{Street23_LSST}
{Street}, R.~A., {Gough-Kelly}, S., {Lam}, C., {et~al.} 2023, arXiv e-prints, arXiv:2306.13792, \dodoi{10.48550/arXiv.2306.13792}

\bibitem[{{Udalski} {et~al.}(1994{\natexlab{a}}){Udalski}, {Szymanski}, {Kaluzny}, {Kubiak}, {Mateo}, \& {Krzeminski}}]{Udalski1994}
{Udalski}, A., {Szymanski}, M., {Kaluzny}, J., {et~al.} 1994{\natexlab{a}}, \apjl, 426, L69, \dodoi{10.1086/187342}

\bibitem[{{Udalski} {et~al.}(1994{\natexlab{b}}){Udalski}, {Szymanski}, {Kaluzny}, {Kubiak}, {Mateo}, {Krzeminski}, \& {Paczynski}}]{Udalski94_EWS}
---. 1994{\natexlab{b}}, \actaa, 44, 227, \dodoi{10.48550/arXiv.astro-ph/9408026}

\bibitem[{{Udalski} {et~al.}(2005){Udalski}, {Jaroszy{\'n}ski}, {Paczy{\'n}ski}, {Kubiak}, {Szyma{\'n}ski}, {Soszy{\'n}ski}, {Pietrzy{\'n}ski}, {Ulaczyk}, {Szewczyk}, {Wyrzykowski}, {Christie}, {DePoy}, {Dong}, {Gal-Yam}, {Gaudi}, {Gould}, {Han}, {L{\'e}pine}, {McCormick}, {Park}, {Pogge}, {Bennett}, {Bond}, {Muraki}, {Tristram}, {Yock}, {Beaulieu}, {Bramich}, {Dieters}, {Greenhill}, {Hill}, {Horne}, \& {Kubas}}]{Udalski05}
{Udalski}, A., {Jaroszy{\'n}ski}, M., {Paczy{\'n}ski}, B., {et~al.} 2005, \apjl, 628, L109, \dodoi{10.1086/432795}

\bibitem[{{Wo\'{z}niak}(2000)}]{Wozniak00}
{Wo\'{z}niak}, P.~R. 2000, Acta Astronomica, 50, 421

\bibitem[{{Wyrzykowski} {et~al.}(2015){Wyrzykowski}, {Rynkiewicz}, {Skowron}, {Koz{\l}owski}, {Udalski}, {Szyma{\'n}ski}, {Kubiak}, {Soszy{\'n}ski}, {Pietrzy{\'n}ski}, {Poleski}, {Pietrukowicz}, \& {Pawlak}}]{Wyrzykowski15_OGLEIII}
{Wyrzykowski}, {\L}., {Rynkiewicz}, A.~E., {Skowron}, J., {et~al.} 2015, \apjs, 216, 12, \dodoi{10.1088/0067-0049/216/1/12}

\bibitem[{Wyrzykowski {et~al.}(2016)Wyrzykowski, Kostrzewa-Rutkowska, Skowron, Rybicki, Mróz, Kozłowski, Udalski, Szymański, Pietrzyński, Soszyński, Ulaczyk, Pietrukowicz, Poleski, Pawlak, Iłkiewicz, \& Rattenbury}]{wyrzykowski2016}
Wyrzykowski, L., Kostrzewa-Rutkowska, Z., Skowron, J., {et~al.} 2016, Monthly Notices of the Royal Astronomical Society, 458, 3012, \dodoi{10.1093/mnras/stw426}

\bibitem[{{Wyrzykowski} {et~al.}(2020){Wyrzykowski}, {Mr{\'o}z}, {Rybicki}, {Gromadzki}, {Ko{\l}aczkowski}, {Zieli{\'n}ski}, {Zieli{\'n}ski}, {Britavskiy}, {Gomboc}, {Sokolovsky}, {Hodgkin}, {Abe}, {Aldi}, {AlMannaei}, {Altavilla}, {Al Qasim}, {Anupama}, {Awiphan}, {Bachelet}, {Bak{\i}{\c{s}}}, {Baker}, {Bartlett}, {Bendjoya}, {Benson}, {Bikmaev}, {Birenbaum}, {Blagorodnova}, {Blanco-Cuaresma}, {Boeva}, {Bonanos}, {Bozza}, {Bramich}, {Bruni}, {Burenin}, {Burgaz}, {Butterley}, {Caines}, {Caton}, {Calchi Novati}, {Carrasco}, {Cassan}, {{\v{C}}epas}, {Cropper}, {Chru{\'s}li{\'n}ska}, {Clementini}, {Clerici}, {Conti}, {Conti}, {Cross}, {Cusano}, {Damljanovic}, {Dapergolas}, {D'Ago}, {de Bruijne}, {Dennefeld}, {Dhillon}, {Dominik}, {Dziedzic}, {Erece}, {Eselevich}, {Esenoglu}, {Eyer}, {Figuera Jaimes}, {Fossey}, {Galeev}, {Grebenev}, {Gupta}, {Gutaev}, {Hallakoun}, {Hamanowicz}, {Han}, {Handzlik}, {Haislip}, {Hanlon}, {Hardy}, {Harrison}, {van Heerden}, {Hoette}, {Horne}, {Hudec}, {Hundertmark}, {Ihanec},
  {Irtuganov}, {Itoh}, {Iwanek}, {Jovanovic}, {Janulis}, {Jel{\'\i}nek}, {Jensen}, {Kaczmarek}, {Katz}, {Khamitov}, {Kilic}, {Klencki}, {Kolb}, {Kopacki}, {Kouprianov}, {Kruszy{\'n}ska}, {Kurowski}, {Latev}, {Lee}, {Leonini}, {Leto}, {Lewis}, {Li}, {Liakos}, {Littlefair}, {Lu}, {Manser}, {Mao}, {Maoz}, {Martin-Carrillo}, {Marais}, {Maskoli{\={u}}nas}, {Maund}, {Meintjes}, {Melnikov}, {Ment}, {Miko{\l}ajczyk}, {Morrell}, {Mowlavi}, {Mo{\'z}dzierski}, {Murphy}, {Nazarov}, {Netzel}, {Nesci}, {Ngeow}, {Norton}, {Ofek}, {Pak{\v{s}}tien{\.{e}}}, {Palaversa}, {Pandey}, {Paraskeva}, {Pawlak}, {Penny}, {Penprase}, {Piascik}, {Prieto}, {Qvam}, {Ranc}, {Rebassa-Mansergas}, {Reichart}, {Reig}, {Rhodes}, {Rivet}, {Rixon}, {Roberts}, {Rosi}, {Russell}, {Zanmar Sanchez}, {Scarpetta}, {Seabroke}, {Shappee}, {Schmidt}, {Shvartzvald}, {Sitek}, {Skowron}, {{\'S}niegowska}, {Snodgrass}, {Soares}, {van Soelen}, {Spetsieri}, {Stankevi{\v{c}}i{\={u}}t{\.{e}}}, {Steele}, {Street}, {Strobl}, {Strubble}, {Szegedi}, {Tinjaca Ramirez},
  {Tomasella}, {Tsapras}, {Vernet}, {Villanueva}, {Vince}, {Wambsganss}, {van der Westhuizen}, {Wiersema}, {Wium}, {Wilson}, {Yoldas}, {Zhuchkov}, {Zhukov}, {Zdanavi{\v{c}}ius}, {Zo{\l}a}, \& {Zubareva}}]{Wyrzykowski20_Gaia16aye}
{Wyrzykowski}, {\L}., {Mr{\'o}z}, P., {Rybicki}, K.~A., {et~al.} 2020, \aap, 633, A98, \dodoi{10.1051/0004-6361/201935097}

\bibitem[{Xu {et~al.}(2020)Xu, Dan, Khim, \& Ravikumar}]{xu2020class}
Xu, Z., Dan, C., Khim, J., \& Ravikumar, P. 2020, in International conference on machine learning, PMLR, 10544--10554

\bibitem[{{Yee} {et~al.}(2021){Yee}, {Zang}, {Udalski}, {Ryu}, {Green}, {Hennerley}, {Marmont}, {Sumi}, {Mao}, {Gromadzki}, {Mr{\'o}z}, {Skowron}, {Poleski}, {Szyma{\'n}ski}, {Soszy{\'n}ski}, {Pietrukowicz}, {Koz{\l}owski}, {Ulaczyk}, {Rybicki}, {Iwanek}, {Wrona}, {Albrow}, {Chung}, {Gould}, {Han}, {Hwang}, {Jung}, {Kim}, {Shin}, {Shvartzvald}, {Cha}, {Kim}, {Kim}, {Lee}, {Lee}, {Lee}, {Park}, {Pogge}, {Bachelet}, {Christie}, {Hundertmark}, {Maoz}, {McCormick}, {Natusch}, {Penny}, {Street}, {Tsapras}, {Beichman}, {Bryden}, {Calchi Novati}, {Carey}, {Gaudi}, {Henderson}, {Johnson}, {Zhu}, {Bond}, {Abe}, {Barry}, {Bennett}, {Bhattacharya}, {Donachie}, {Fujii}, {Fukui}, {Hirao}, {Ishitani Silva}, {Itow}, {Kirikawa}, {Kondo}, {Koshimoto}, {Li}, {Matsubara}, {Muraki}, {Miyazaki}, {Olmschenk}, {Ranc}, {Rattenbury}, {Satoh}, {Shoji}, {Suzuki}, {Tanaka}, {Tristram}, {Yamawaki}, \& {Yonehara}}]{Yee21_ob0960}
{Yee}, J.~C., {Zang}, W., {Udalski}, A., {et~al.} 2021, arXiv e-prints, arXiv:2101.04696.
\newblock \doarXiv{2101.04696}

\bibitem[{{Zang} {et~al.}(2020){Zang}, {Dong}, {Gould}, {Calchi Novati}, {Chen}, {Yang}, {Li}, {Mao}, {Alton}, {Brimacombe}, {Carey}, {Christie}, {Delplancke-Str{\"o}bele}, {Feliz}, {Gaudi}, {Green}, {Hu}, {Jayasinghe}, {Koff}, {Kurtenkov}, {M{\'e}rand }, {Minev}, {Mutel}, {Natusch}, {Roth}, {Shvartzvald}, {Sun}, {Vanmunster}, \& {Zhu}}]{Zang20Kojima}
{Zang}, W., {Dong}, S., {Gould}, A., {et~al.} 2020, \apj, 897, 180, \dodoi{10.3847/1538-4357/ab9749}

\bibitem[{{Zang} {et~al.}(2021{\natexlab{a}}){Zang}, {Hwang}, {Udalski}, {Wang}, {Zhu}, {Sumi}, {Yee}, {Gould}, {Mao}, {Zhang}, {Albrow}, {Chung}, {Han}, {Jung}, {Ryu}, {Shin}, {Shvartzvald}, {Cha}, {Kim}, {Kim}, {Kim}, {Lee}, {Lee}, {Lee}, {Park}, {Pogge}, {Mr{\'o}z}, {Skowron}, {Poleski}, {Szyma{\'n}ski}, {Soszy{\'n}ski}, {Pietrukowicz}, {Koz{\l}owski}, {Ulaczyk}, {Rybicki}, {Iwanek}, {Wrona}, {Gromadzki}, {Bond}, {Abe}, {Barry}, {Bennett}, {Bhattacharya}, {Donachie}, {Fujii}, {Fukui}, {Hirao}, {Itow}, {Kirikawa}, {Kondo}, {Koshimoto}, {Li}, {Matsubara}, {Muraki}, {Miyazaki}, {Olmschenk}, {Ranc}, {Rattenbury}, {Satoh}, {Shoji}, {Ishitani Silva}, {Suzuki}, {Tanaka}, {Tristram}, {Yamawaki}, {Yonehara}, {Beichman}, {Bryden}, {Calchi Novati}, {Carey}, {Gaudi}, {Henderson}, {Johnson}, \& {Spitzer Team}}]{Zang21AF1}
{Zang}, W., {Hwang}, K.-H., {Udalski}, A., {et~al.} 2021{\natexlab{a}}, \aj, 162, 163, \dodoi{10.3847/1538-3881/ac12d4}

\bibitem[{{Zang} {et~al.}(2021{\natexlab{b}}){Zang}, {Han}, {Kondo}, {Yee}, {Lee}, {Gould}, {Mao}, {de Almeida}, {Shvartzvald}, {Zhang}, {Albrow}, {Chung}, {Hwang}, {Jung}, {Ryu}, {Shin}, {Cha}, {Kim}, {Kim}, {Kim}, {Lee}, {Lee}, {Park}, {Pogge}, {Drummond}, {Tan}, {Dias do Nascimento J{\'u}nior}, {Maoz}, {Penny}, {Zhu}, {Bond}, {Abe}, {Barry}, {Bennett}, {Bhattacharya}, {Donachie}, {Fujii}, {Fukui}, {Hirao}, {Itow}, {Kirikawa}, {Koshimoto}, {Li}, {Matsubara}, {Muraki}, {Miyazaki}, {Ranc}, {Rattenbury}, {Satoh}, {Shoji}, {Sumi}, {Suzuki}, {Tanaka}, {Tristram}, {Yamawaki}, {Yonehara}, {Petric}, {Burdullis}, \& {Fouqu{\'e}}}]{Zang21_kb0414}
{Zang}, W., {Han}, C., {Kondo}, I., {et~al.} 2021{\natexlab{b}}, arXiv e-prints, arXiv:2103.01896.
\newblock \doarXiv{2103.01896}

\end{thebibliography}
\bibliographystyle{aasjournal}

\appendix
\section{Confusion Matrices}\label{sec:confmat}

\begin{figure}[h!]
    \centering
    \includegraphics[width=\columnwidth]{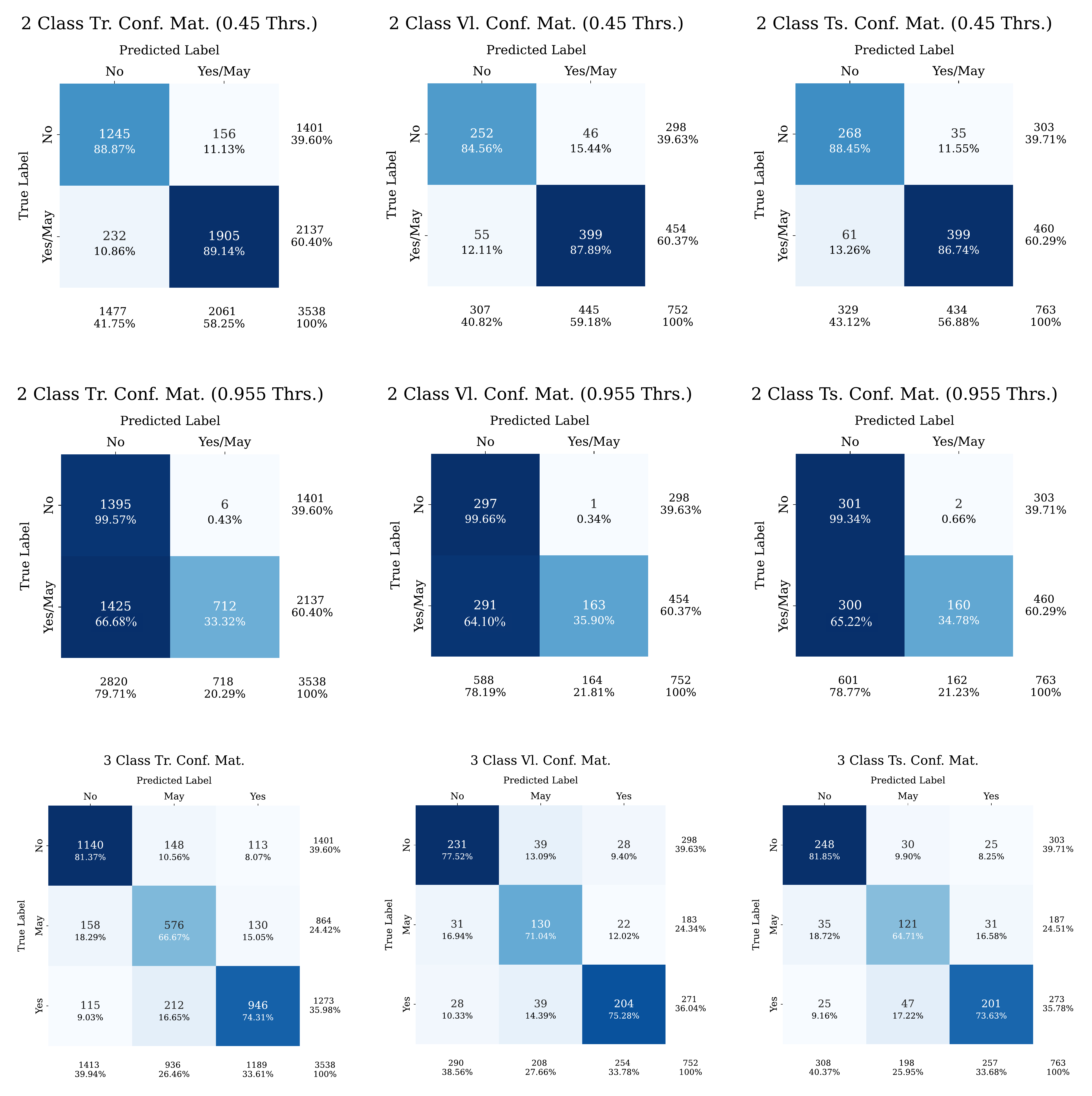}
    \caption{Confusion matrices of LensNet calculated using the non-augmented data, showing the performance in the binary classification task (with the two threshold values discussed of 0.45 and 0.955) and in the 3-class classification task. The accuracy rates for both tasks are displayed on the matrices for each set (training, validation and testing). These results highlight the model's overall performance.}
    \label{fig:fig_13_conf_mats}
\end{figure}

\end{document}